\documentclass[11pt,a4paper]{article}
\usepackage{jheppub}

\usepackage[utf8]{inputenc}
\usepackage{amsmath}
\usepackage{amsfonts,dsfont,mathrsfs}
\usepackage{blkarray}
\usepackage{subcaption}
\usepackage[colorlinks=true]{hyperref}

\usepackage{mathtools}
\newcommand{\eqdef}{=\vcentcolon}
\def\rt#1{\langle #1 \rangle}

\usepackage{tikz}
\usetikzlibrary{decorations.pathreplacing}
\usetikzlibrary{decorations.pathmorphing}
\usetikzlibrary{decorations.markings}
\usetikzlibrary{arrows}
\usetikzlibrary{shapes}
\usetikzlibrary{matrix}
\usepackage[english]{babel}
\usepackage[autostyle]{csquotes}
\usepackage{braket}

\tikzset{cross/.style={cross out, draw=black}}
\tikzset{circ/.style={circle,fill=white,draw=black}}
\tikzset{hasse/.style={circle, fill,inner sep=2pt}}
\tikzset{h/.style={circle, fill,inner sep=2pt}}
\tikzset{ns/.style={circle, draw,inner sep=2pt}}
\tikzset{gauge/.style={circle,draw}}
\tikzset{flavor/.style={regular polygon,regular polygon sides=4, draw}}
\tikzset{triplearrow/.style={
  draw=black!75,
  color=black!75,
  thick,
  double distance=3pt, 
}}
\tikzset{thirdline/.style={
  draw=black!75,
  color=black!75,
  thick,
  <-,
}}
\tikzset{middlearrow/.style={
        decoration={markings,
            mark= at position 0.5 with {\arrow{#1}} ,
        },
        postaction={decorate}
    }
}
\newcommand{\midarrow}{\tikz \draw[-triangle 90] (0,0) -- +(.1,0);}
\newcommand{\midarrowrev}{\tikz \draw[-triangle 90]  +(.1,0) -- (0,0);}

\newcommand{\C}{\mathcal C}
\newcommand{\higgs}{\mathcal H}
\newcommand{\co}{\mathbb C}

\newcommand{\m}{\mathbf m}

\def\v#1{V_{\langle #1 \rangle}}
\def\p#1{\varphi_{#1}}

\def\m#1{M_{#1}}
\def\up#1{u^+_{#1}}
\def\um#1{u^-_{#1}}
\def\upm#1{u^\pm_{#1}}
\def\bmat#1{\left(\begin{array}{#1}}
\def\emat{\end{array}\right)}
\def\be{\begin{equation}} \def\ee{\end{equation}}
\def\ba{\begin{align}} \def\ea{\end{align}}
\def\sl#1{\mathfrak{sl}(#1,\mathbb{C})}
\def\so#1{\mathfrak{so}(#1,\mathbb{C})}

\newcommand\pois[1]{\lbrace #1 \rbrace}
\newcommand\tr[1]{\mathrm{Tr}(#1)}
\newcommand{\im}{\mathrm{i}}

\allowdisplaybreaks

\title{Nilpotent orbit Coulomb branches of types $AD$}
\author[a]{Amihay Hanany}
\author[b]{Dominik Miketa}
\affiliation[a,b]{Theoretical Physics, The Blackett Laboratory\\
Imperial College London\\ SW7 2AZ United Kingdom}

\emailAdd{a.hanany@imperial.ac.uk, d.miketa16@imperial.ac.uk}

\abstract{We develop a new method for constructing $3d$ $\mathcal{N}=4$ Coulomb branch chiral rings in terms of gauge-invariant generators and relations while making the global symmetry manifest. Our examples generalise to all balanced quivers of type $A$ and $D$ whose Coulomb branches are closures of nilpotent orbits. This new approach is a synthesis of operator counting using Hilbert series and explicit algebraic construction introduced by Bullimore, Dimofte and Gaiotto with significant potential for further generalisation to other quivers, including non-simply laced. The method also identifies complex mass deformations of many Coulomb branches, providing an explicit construction for complex deformations of nilpotent orbits.}


\begin{document}


\maketitle
\flushbottom
\section{Introduction}

The space of admissible vacua, or the moduli space, is among the simplest characteristics of a quantum field theory. It is parametrised by vacuum expectation values of gauge-invariant operators transforming as scalars under the Poincaré group. Despite their simplicity, moduli spaces can be highly structured and mathematically interesting objects. Theories with 8 supercharges present a particularly rich selection of interesting and significant examples: they naturally occur as moduli spaces of brane systems in string theory, we have good control over them thanks to supersymmetry, they feature genuinely interesting and calculable non-perturbative physics and can be connected by a web of dualities to moduli spaces of other theories with 8 supercharges. For example, it was recently discovered that three-dimensional $\mathcal{N}=4$ theories hold a wealth of interesting information about five-dimensional $\mathcal{N}=1$ gauge theories \cite{Ferlito_3d_2017}.

One often restricts to the part of moduli space which is parametrised by the set of (simultaneously) $1/2$-BPS operators -- the chiral ring -- on account of improved theoretical control. It is then possible to naturally break the moduli space into several qualitatively distinct subspaces, or branches. The Higgs branch is a Hyperk\"{a}hler manifold and a supersymmetric non-renormalisation theorem protects it from quantum corrections; it is also essentially the same in every dimension\footnote{\label{footnote:higgsbr}With a caveat: the Higgs branch is quantum-mechanically corrected in 5 and 6 dimensions, but only in the infinite coupling regime \cite{Cremonesi_Instanton_2017,Ferlito_3d_2017}, and in 4 dimensions at Argyres-Douglas points \cite{ARGYRES199593}.}. Coulomb branches at classical and fully quantum field theoretical levels can differ greatly, however. This is observed most dramatically in three dimensions where new inherently non-perturbative particles -- topological vortices -- emerge in the deep IR and open new directions in the moduli space. The new space is also Hyperk\"{a}hler and often exhibits highly non-trivial isometries.

Coulomb branches of $3d$ $\mathcal{N}=4$ theories have received much attention since the mid-nineties and the following summary is by no means exhaustive. Initial investigations exploited mirror symmetry which relates a Coulomb branch of one theory to a Higgs branch of another theory \cite{Intriligator_Mirror_1996,HANANY1997152}. Later papers treated the theory in an SCFT framework and produced relatively limited but valuable results for some key examples \cite{Borokhov_Topological_2003,Borokhov_Monopole_2002}. The authors of \cite{Gaiotto_S_2008} were able to generalise them for a wide range of three-dimensional quivers and laid the groundwork for operator counting \cite{Cremonesi:2013lqa,Hanany_Algebraic_2016,Hanany:2016gbz,Hanany:2017ooe} . This work made it clear that Coulomb branches provide interesting new examples of Hyperk\"{a}hler varieties and more mathematicians had become interested in this topic as a result \cite{Braverman_Coulomb_2016,Braverman_Towards_2016}. Finally, several recent papers have provided algebraic constructions of the Coulomb branch chiral ring, albeit limited in scope to unitary nodes and linear quivers \cite{Bullimore_The_2017,Bullimore_Vortices_2016,Assel_Ring_2017,Assel:2017jgo}, or a single symplectic node \cite{Assel:2018exy}.

Unsurprisingly, each approach comes with its strengths and drawbacks. Operator counting is very general, straightforwardly algorithmic and naturally captures the isometry of the Coulomb branch and representation-theoretic content of chiral ring relations, reducing the problem of finding the moduli space to identifying coefficients for finitely many linear combinations of finitely many operators. The representation-theoretic data is also often sufficient to solve this latter problem. However, turning on  complex mass deformations compromises the computational utility of this method. Operator counting also rarely aids physical interpretation of particular chiral ring operators. On the other hand the recent algebraic construction leverages operators' physical properties, naturally handles complex mass deformations and in principle fully specifies the moduli space for arbitrary quivers. However, the way in which it is defined obscures the isometry, corresponding representation-theoretic data and as a result physical relations between gauge-invariant chiral operators are difficult to extract.

This paper demonstrates that operator counting and algebraic construction can be synthesised into a new method which combines their strengths, removes many of their drawbacks and provides a new and powerful way to derive relations between gauge-invariant operators in $3d$ $\mathcal{N}=4$ theories. Our examples are drawn from families of balanced quiver gauge theories of type $A$ and $D$ (ie. shaped like their namesakes among Dynkin diagrams) and, in the case of $D$, of Panyushev height\footnote{Panyushev height is defined in Sec. \ref{sec:quiver}.} 2; these examples have been studied in \cite{Hanany:2016gbz} using only operator counting. A follow-up paper will expand this work to types $B$ and $C$, also of height 2. 

In Sec. \ref{sec:general} we provide a brief introduction into generalities of quivers, their Coulomb branch chiral rings and operator counting. Sec. \ref{sec:Atype} develops our tools for quivers of type $A$, largely following the pioneering work of \cite{Bullimore_The_2017}. Sec. \ref{sec:Dtype} expands the method to quivers of type $D$ and height 2. We close with Sec. \ref{sec:future} sketching out the wide variety of directions that are now open to investigation.

\section{General remarks} \label{sec:general}

\subsection{Quivers} \label{sec:quiver}

\begin{figure}
	\begin{subfigure}[t]{0.17\textwidth}
	\centering
		\begin{tikzpicture}
			\node (g1) [gauge, label=below:{\textbf{1}}]{};
			\node (g2) [gauge, right of=g1, label=below:{\textbf{2}}]{};
			\node (g3) [gauge, right of=g2, label=below:{\textbf{3}}]{};
			\node (f3) [flavor, above of=g3,label=above:{\textbf{4}}]{};
			\draw (g1)--(g2)--(g3);
        	\draw (g3)--(f3);
			\end{tikzpicture}
	\end{subfigure}
~
	\begin{subfigure}[t]{0.22\textwidth}
	\centering
	\begin{tikzpicture}
		\node (g1) [gauge, label=below:{\textbf{1}}]{};
		\node (g2) [gauge, right of=g1, label=below:{\textbf{2}}]{};
		\node (gn-2) [gauge, right of=g2, label=below:{\textbf{2}}]{};
		\node (gn-1) [gauge, right of=gn-2, label=below:{\textbf{1}}]{};
		\node (f2) [flavor, above of=g2,label=above:{\textbf{1}}]{};
		\draw (g1)--(g2)--(gn-2);
		\draw[triplearrow] (gn-2)-- node {\midarrow} (gn-1);
        \draw (g2)--(f2);
		\end{tikzpicture}	
	\end{subfigure}
~	
	\begin{subfigure}[t]{0.22\textwidth}
	\centering
		\begin{tikzpicture}
		\node (g1) [gauge, label=below:{\textbf{1}}]{};
		\node (g2) [gauge, right of=g1, label=below:{\textbf{1}}]{};
		\node (g3) [gauge, right of=g2, label=below:{\textbf{1}}]{};
		\node (f1) [flavor, above of=g1,label=above:{\textbf{1}}]{};
		\draw (g1)--(g2);
        \draw (g1)--(f1);
        \draw[triplearrow] (g3)-- node {\midarrowrev} (g2);
		\end{tikzpicture}
	\end{subfigure}
~
	\begin{subfigure}[t]{0.22\textwidth}
	\centering
	\begin{tikzpicture}
		\node (g1) [gauge, label=below:{\textbf{1}}]{};
		\node (g2) [gauge, right of=g1, label=below:{\textbf{2}}]{};
		\node (g3) [gauge, right of=g2, label=below:{\textbf{2}}]{};
		\node (gn-2) [gauge, right of=g3, label=below:{\textbf{2}}]{};
		\node (gn-1) [gauge, above right of=gn-2, label=below:{\textbf{1}}]{};
		\node (gn) [gauge, below right of=gn-2, label=below:{\textbf{1}}]{};
		\node (f2) [flavor, above of=g2,label=above:{\textbf{1}}]{};
		\draw (g1)--(g2)--(g3)--(gn-2);
		\draw (gn-2)--(gn-1);
		\draw (gn-2)--(gn);
        \draw (g2)--(f2);
		\end{tikzpicture}	
	\end{subfigure}
	\caption{Examples of balanced quivers of type $A$, $B$, $C$, resp. $D$}
	\label{fig:quiver-examples}
\end{figure}
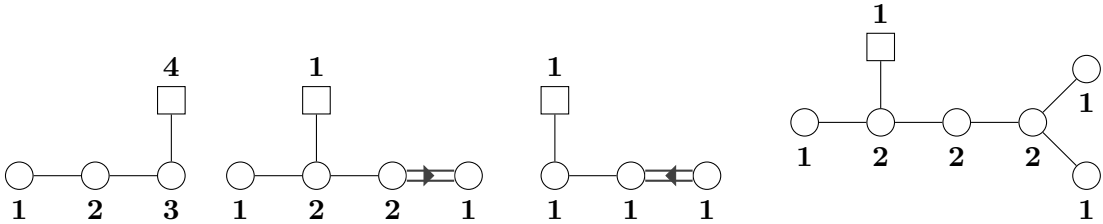

We investigate $\mathcal{N}=4$ balanced simply laced quiver gauge theories in $2+1$ dimensions with unitary gauge groups. Such theories are specified by a connected\footnote{We restrict to connected graphs since unconnected graphs describe decoupled sectors and hence add nothing new to the discussion.} graph called a quiver. Its circular nodes signify unitary\footnote{There are also orthosymplectic quiver gauge theories whose gauge nodes alternate between orthogonal and symplectic groups, but we do not consider them here.} groups $U(r)$, whose product forms the overall gauge group. Each gauge factor comes with supersymmetric vector multiplets in the adjoint representation while each undirected link between two nodes corresponds to a hypermultiplet transforming under the fundamental (or anti-fundamental) representation under both nodes connected by the link\footnote{This is the only type of link in this paper although others exist: \cite{Cremonesi:2014xha} study ``multiple" and directed links suggestively reminiscent of Dynkin diagrams of types $B$, $C$, $F$ and $G$. All quivers in this paper are of type $A$ or $D$, hence ``simply laced".}. We will only consider links which start and end on different nodes, ie. our theories will not include adjoint hypermultiplets. Each circular node can be connected to a square node representing flavor symmetry. Links connecting a square node to a circular node describe a matter hypermultiplet charged under the fundamental representation of both the gauge group and the flavor group. Several simple quivers are presented in Fig. \ref{fig:quiver-examples}.

These and similar quiver descriptions straightforwardly prescribe a Lagrangian and vice versa\footnote{With the possible exception of non-simply laced quivers whose Lagrangians are unknown. We have recently made some progress in developing a well-defined Lagrangian description and hope to present it in a followup paper.}. In the absence of a flavor node an overall $U(1)$ gauge subgroup decouples and can effectively provide a flavor node of rank 1. We preempt this decoupling phenomenon by only considering quivers with at least one flavor node.

In the interest of further simplicity, all gauge nodes are required to be \emph{balanced}. This is easily defined: start with the rank of a gauge node $U(r)$, which is just $r$. Then let $n$ be the number of hypermultiplets charged under it; this will be the sum of ranks of all adjacent gauge and flavor groups. Then, following \cite{Gaiotto_S_2008}, we define \emph{excess} as

\be
e=n-2r
\ee

If $e=0$ we say the node is balanced. If the excess is zero at every gauge node then the quiver as a whole is said to be balanced. All quivers in this paper are balanced but we intend to adapt our techniques to quivers with unbalanced nodes in future work\footnote{\cite{Dimofte:2018abu} treats a class of star-shaped quivers, which generally involve unbalanced nodes, with similar tools.}


Finally, all quivers in this paper fall into two categories: either their gauge subgraph looks like the Dynkin diagram for $\sl{n}$, in which case we say the quiver is of type $A$, or the gauge subgraph is that of $\so{2n}$ and the quiver is of type $D$. We will be able to cover all quivers of type $A$ (subject to stated restrictions) but have to impose a final condition on type $D$: such quivers must have height 2.

The definition of quiver height \cite{PANYUSHEV} leverages the similarity between Dynkin diagrams of simple Lie algebras and subgraphs of quivers formed by all gauge nodes and can be calculated by taking a dot product between the vector of Coxeter labels and the vector of flavor ranks. For example, the family of type $D$ quivers depicted in Fig. \ref{fig:Dmin}, whose Coxeter labels are $(1,2,\dots,2,1,1)$, has height $(0,1,\dots,0,0,0)\cdot(1,2,\dots,2,1,1)= 2$.

Although it may seem that we have narrowed the class of quivers almost out of existence, we have merely restricted to cases covered in \cite{Hanany:2016gbz}, whose Coulomb branches are closures of \emph{nilpotent orbits}\footnote{We will trade accuracy for brevity and refer to closures of nilpotent orbits as, simply, ``nilpotent orbits" in what follows.}. They are the simplest exemplars of their kind and hence a suitable arena for development of a new technique. We expect that once our method is established for these basic cases most -- if not all -- of the imposed restrictions can be lifted and the description will generalise to varieties beyond nilpotent orbits.

\subsection{Chiral ring} \label{sec:chiralring}

Among the simplest aspects of a quiver theory one can study is its \emph{moduli space}, or the set of all admissible vacuum expectation values of Lorentz-invariant operators. We restrict our attention to \emph{chiral} operators which break one half of $\mathcal{N}=4$ supersymmetry. Let $\mathcal{O}_1(x)$ and $\mathcal{O}_2(y)$ be two such operators. General results show that $\mathcal{O}_1(x) \mathcal{O}_2(y)$ is also chiral and furthermore $\langle \mathcal{O}_1(x) \mathcal{O}_2(y)\rangle$ independent of $x$ or $y$ and we can suppress them. Moreover cluster decomposition implies that $\langle \mathcal{O}_1 \mathcal{O}_2 \rangle = \langle \mathcal{O}_1 \rangle \langle \mathcal{O}_2 \rangle$. It follows that vevs of chiral operators $\langle \mathcal{O}_i \rangle$ form a ring. One can treat it as a coordinate ring and attempt to reconstruct and study the corresponding algebraic variety formed of all possible vacua. That is the general motivation of this paper.


Classical $F$-term equations imply that a non-zero vev on one chiral operator may impose a zero vev condition on other operators and supersymmetric non-renormalisation theorems (which naturally take into account $D$-terms) show that this feature persists in the quantum theory. In this way the moduli space breaks into several branches: the Higgs branch $\higgs$, the Coulomb branch $\C$ and a number of mixed branches. The Higgs branch admits vevs on all scalar operators originating from matter hypermultiplets and it is protected from quantum corrections by supersymmetry\footnote{See footnote \ref{footnote:higgsbr}.}. It is mathematically interesting in its own right as a concrete example of a Hyperk\"{a}hler quotient. Mixed branches are essentially combinations of Higgs and Coulomb branches and while noteworthy we will not consider them or the Higgs branch in this paper. Instead our focus will be on the Coulomb branch. It can be morally defined as the subset of vacua where no scalars in matter hypermultiplets exhibit vevs but scalars in vector multiplets possibly do. The chiral ring is then precisely the coordinate ring of $\C$ and we will denote it $\co\lbrack \C \rbrack$. 

We are interested in the vacuum manifold so it is natural to consider the theory in the deep IR. The only chiral operators with (potentially) non-zero vevs are the gauge-invariant combinations of scalar superpartners of gauge bosons, which are present in the theory's Lagrangian, and monopole operators, which become relevant in the deep IR. Loosely speaking, monopole operators serve as creation and annihilation operators for topological particles called \emph{vortices}. Turning on monopole operator vevs leaves $F-$ and $D$-terms (and hence the conditions for a vacuum) intact. Consequently both kinds of operators (and their products) admit simultaneous non-zero vevs. 


The chiral ring can be presented as a freely generated ring quotiented by an ideal:

\be
\co\lbrack\C\rbrack = \co\lbrack \mathcal{O}_1,\mathcal{O}_2,\dots\rbrack /  \mathcal{I} \label{eq:chiralring}
\ee

We will refer to the $\mathcal{O}_i$ -- which stand in for vevs of gauge-invariant chiral operators -- as \emph{generators}. Elements of $\mathcal{I}$ are called \emph{relations} and we usually find that the ideal is non-trivial but finitely generated.

Presence of flavor nodes indicates hypermultiplets in the Lagrangian which can be given complex mass by conventional means without breaking more supersymmetry. Quiver theories are often studied at the IR superconformal point where all masses are set to zero, but once the SCFT is understood one can turn on real and complex mass parameters and study how its moduli space deforms. Our method is particularly suited for investigations of complex mass deformations. \footnote{Real mass deformations were recently studied in \cite{Hanany:2018uzt}. It would be interesting to integrate them with methods covered in this paper.}

Coulomb branches with unitary nodes always exhibit some isometry and generators will assemble into irreducible representations of this symmetry's Lie algebra. We can be even more precise with the subset of quivers to which we restrict: chiral ring generators form the adjoint (or coadjoint) representation of the overall isometry, which can in turn be read off the quiver reinterpreted as a Dynkin diagram for a simple Lie algebra. So quivers of type $A$ with $n$ nodes have $\sl{n+1}$ isometry on their Coulomb branch while quivers of type $D$ with the same number of nodes exhibit $\so{2n}$ isometry \cite{Gaiotto_S_2008} and the entire chiral ring is generated by (components of) the appropriate (co)adjoint representation \cite{Hanany:2016gbz}. 

Relations can be stated in the form of various contractions of the adjoint tensor or conditions on minors thereof. This is a very desirable presentation of the chiral ring because it makes the isometry of $\C$ explicit. It also allows direct contact with a family of well understood spaces: the \emph{nilpotent orbits} of Lie algebras\footnote{Roughly speaking, nilpotent orbits are well-defined subspaces of upper-triangular matrices in a fixed Lie algebra, invariant under its adjoint action.} \cite{Hanany:2016gbz}. We will say more when we are able to get the generators and relations in this form. But for that we need to first introduce operator counting and algebraic construction of the chiral ring, which we then combine into a ``synthetic" method for determining the full gauge-invariant presentation of the chiral ring.

\subsection{Operator counting and monopole operators}

We will now briefly review operator counting, or the Hilbert series approach to Coulomb branch chiral rings. For more background see \cite{Cremonesi:2013lqa,Cremonesi_3d_2017,Hanany:2016gbz}. The two main insights behind this method are that we can often easily identify a set of ``basic" symmetries of the Coulomb branch and that we in principle know exactly how many operators carry any particular combination of charges under these symmetries. This information is preserved by ring isomorphisms, so it in particular has to be the same for any description of the physical chiral ring (which we can specify) and the coordinate ring of a putative geometric description of the Coulomb branch (which we would like to find) and constitutes a highly non-trivial test which is sometimes sufficient to fully specify the presentation in \ref{eq:chiralring}.

Monopole operators have a ready path integral interpretation as disorder operators which insert a Dirac singularity into the gauge field \cite{Borokhov_Topological_2003}. Three dimensional magnetic monopole operators are local operators but they are still charged under the dual (or Langlands) gauge group $\check{G}\simeq\prod_i U(r_i)$ \cite{GNO}. Specifically, the set of admissible magnetic charges forms the principal Weyl chamber of the dual gauge group's weight lattice 

\be
\Gamma_{\check{G}}/\mathcal{W}_{\check{G}} = \prod_i \Gamma_{U(r_i)}/\mathcal{W}_{U(r_i)}.
\ee 

Each $\Gamma_{U(r_i)}/\mathcal{W}_{U(r_i)}$ holds $r_i$ integer-valued magnetic charges $m_{i,j}$ ordered in non-increasing fashion by the action of the Weyl group:

\be
m_{i,1}\geq m_{i,2} \geq \cdots \geq m_{i,r_i}
\ee

Every magnetic monopole carries some non-zero array of magnetic charges while scalar operators are inert under the dual gauge group. A product of monopole and scalar operator then has the same magnetic charges as the original monopole operator; we say the scalar operators \emph{dress} the monopole\footnote{Explicit construction of the chiral ring makes it clear that it is scalar operators, not their gauge-invariant combinations in Casimir operators, which dress monopole operators. In the notation of later sections, $u^+_{2,1}\varphi_{2,2} + u^+_{2,2}\varphi_{2,1}$ is a dressed monopole operator even though it cannot be factorised into monopole and Casimir operators.}. The chiral ring always contains a ``basic" monopole of a particular magnetic charge stripped of any factors of scalar fields, which we call a \emph{bare} monopole operator.

One might hope that monopole operators could be straightforwardly labelled (and hence counted) by their magnetic charges but there is a subtlety which prevents this. It turns out that magnetic charges are not conserved for general gauge groups, which disqualifies them for counting purposes. But they are conserved for $U(1)$ gauge groups -- and each $U(r_i)$ includes a $U(1)$ factor. The Hodge dual of its field strength $J=\star F^{U(1)}$ is a conserved current on account of Bianchi identity $\mathrm{d} F^{U(1)}$=0 and independently of equations of motion:

\be
\star \mathrm{d} \star J = - \star \mathrm{d} F^{U(1)} = 0.
\ee

The conserved current $J$ is called \emph{topological} due to its relation to twists of the gauge group's principal bundle. Any conserved current indicates the presence of a continuous symmetry by Noether's theorem. The topological charge under this symmetry is given by:

\be
q_i = \sum_{j=1}^{r_i} m_{i,j} \in \mathbb{Z}
\ee

Each monopole operator can have any combination of integral topological charges (even 0 at every node) while scalar operators are always topologically uncharged.


There is one final charge to consider. The $R$-symmetry of $3d$ $\mathcal{N}=4$ theories is $SO(4) \simeq SU(2)_C \times SU(2)_H$. The factor $SU(2)_C$ acts on Coulomb branch operators while $SU(2)_H$ acts on operators in the Higgs branch. Both branches are Hyperk\"{a}hler and therefore each carries three complex structures arranged into triplets of the respective $SU(2)$ symmetry. We restrict our attention to the Coulomb branch and so disregard $SU(2)_H$. We choose an arbitrary complex structure on the Coulomb branch, which is equivalent to selecting an $\mathcal{N}=2$ subalgebra or fixing the meaning of ``chiral" by designating unbroken supercharges. $SU(2)_C$ merely rotates this choice of complex structure, $\mathcal{N}=2$ subalgebra or unbroken supercharges. Finally we restrict to holomorphic functions under this complex structure as the rest can be reached by $SU(2)_C$ rotations.

Operators carry a charge under $SU(2)_C$ action; we refer to it as the $R$-symmetry spin and normalise it so that the lowest non-trivial spin is $1/2$, as is common in physics literature. Restriction to holomorphic functions is equivalent to restriction to highest weight representatives within $SU(2)_C$ multiplets. Total $SU(2)_C$ spin of a product of two such operators is therefore just the sum of of their individual spins. We say that spin is additive.

$R$-symmetry spin remains constant for protected operators throughout RG flow into the deep IR \cite{Gaiotto_S_2008} in good and ugly theories (which include all quivers in this paper). Since we are only interested in protected operators we exploit this property to calculate $R$-symmetry spin of an arbitrary Coulomb branch operator using the monopole formula of \cite{Cremonesi:2013lqa}. We cite it in the form adapted to a unitary simply laced quiver. Its gauge nodes are labelled by $i\in I$, the set of nodes adjacent to $i$ is denoted $A(i)$ and the number of attached flavors is $s_i$. The $R$-symmetry spin $\Delta$ of a bare monopole operator is given by:

\be
\Delta(\mathbf{m})= -\sum_{i\in I}\sum_{a=1}^{r_i-1} | m_{i,a}-m_{i,a+1} | + \frac{1}{4} \sum_i \sum_{j\in A(i)} \sum_{a=1}^{r_i} \sum_{b=1}^{r_j} |m_{i,a} - m_{j,b}| + \frac{1}{2} \sum_i \sum_{a=1}^{r_i} s_i |m_{i,a}|
\ee

(The unusual factor of $1/4$ in front of the second term compensates for the fact that we technically sum twice over each link between gauge nodes.)

Scalar operators have $R$-symmetry spin 1, but only enter the chiral ring in gauge-invariant combinations. Since scalars belong to vector supermultiplets along with gauge vectors, they necessarily transform under the adjoint representation of the gauge group just like their vector superpartners. Their gauge-invariant combinations are then precisely the Casimir invariants of the gauge group (and their sums and products). The invariant of lowest order in scalars is merely linear and so its $R$-symmetry spin is also 1. We will henceforth refer to gauge-invariant combinations of scalar operators in vector supermultiplets as Casimir operators and will reserve the term \emph{linear Casimir} for Casimir operators of spin 1.

Since spin is additive we can reconstruct $R$-symmetry spin for any monopole operator by summing up the contribution due to magnetic charges with the contribution of scalar dressing. So $R$-symmetry spins are known for all operators and we can ask how many linearly independent chiral ring operators there are for a given spin $s$. Such operators form a vector space $V_s$, so we are effectively inquiring about $\dim V_s$. This data is typically repackaged as an infinite series

\be
\mathrm{HS}(t) = \sum_{2s=0}^\infty ( \dim V_s )\ t^{s}
\ee

called the (unrefined) Hilbert series of $\C$. It can be naturally ``refined" by bringing in topological charges of each operator:

\be
\mathrm{HS}(t) = \sum_{2s=0}^\infty\sum_{\substack{q_1=-\infty\\ q_2=-\infty\\ \vdots}}^\infty ( \dim V_{s,q_1,q_2,\dots} )\ t^{s} \prod_i z_i^{q_i}
\ee
where $V_{s,q_1,q_2,\dots}$ is the vector space of all chiral ring operators with $R$-symmetry spin $s$ and topological charges $\vec{q}=\rt{q_1,q_2,\dots}$. The $z_i$ are called \emph{topological fugacities}.

Now comes the crucial part: the polynomial multiplying $t^s$ is -- trivially -- a character of the topological symmetry $\prod_i U(1)$. But it may also be a character for a \emph{larger} group. That could happen by chance for a particular order in $t$, but it would be much less likely that \emph{all} coefficients of $t^s$, for all $s$, are characters of the same larger group. Such a coincidence provides strong evidence of \emph{enhanced symmetry} of the Coulomb branch. For example the Coulomb branch of the quiver in Fig. \ref{fig:SQED} has topological symmetry $U(1)$ coming from its single gauge node, but the coefficient of $t$ at every order in the Hilbert series is an $SU(2)$ character. The Coulomb branch isometry is then likely enhanced to overall $SU(2)$. New directions on the Coulomb branch correspond to vevs of monopole operators; we will shortly see this example worked out in explicit detail.

Note that the Hilbert series is preserved under complex mass deformation. If we read off the isometry of the SCFT Coulomb branch from the Hilbert series, and the series remains untouched upon turning on complex mass parameters, it is natural to conjecture that the isometry will also remain intact. We will be able to confirm it for worked examples.


So Hilbert series suggests the isometry; it also gives us quite a bit more than that. The coefficient at the lowest non-trivial order in $t$ must correspond to (at least some of) the generators. The Casimirs must be linear if they are present at that order at all and the monopole operators must be bare. In fact the rather special quivers in this paper always have Coulomb branch chiral rings generated by operators at order $t$, ie. by linear Casimirs and (specific) bare monopole operators. They assemble into the adjoint representation of the isometry -- and the isometry is, as was already mentioned, precisely the simple Lie algebra represented by the quiver reinterpreted as a quiver diagram. This claim contains a slight subtlety as the Hilbert series does not distinguish between adjoint and coadjoint representations due to their isomorphism, and indeed, we will see that the natural objects to come out of our calculations are coadjoint. At the level of the Hilbert series, however, there is no difference. Note that other types of quivers may have chiral rings generated by operators beyond lowest order in $t$. 

Finally, operator counting can pin down the relations between generators. This is largely thanks to its sensitivity to isometry: if generators form tensors of the isometry then so must relations, since otherwise they would break the symmetry. Close analysis of a calculated Hilbert series will typically reveal that there are fewer operators at higher orders in $t$ than would be expected from free (symmetric) products of generating tensors; they must be ``removed" by a set of relations which transform in irreducible representations of the isometry\footnote{This claim can be recast in more technical terms of plethystic logarithms and syzygies \cite{Benv-Feng-Hanany-He-2007,Feng:2007ur}.}.

\section{Type $A$} \label{sec:Atype}

\subsection{$\sl{2}$: A simple example}

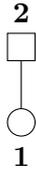
\begin{figure}[t]
	\centering
	\begin{tikzpicture}
		\node (g1) [gauge, label=below:{\textbf{1}}]{};
		\node (f1) [flavor, above of=g1,label=above:{\textbf{2}}]{};
        \draw (g1)--(f1);
		\end{tikzpicture}
		\caption{SQED with 2 electrons}
		\label{fig:SQED}
\end{figure}

The main results of this paper are best introduced as generalisations of two concrete results, both of which originally appeared in \cite{Bullimore_The_2017} in some form. The simpler of the two concerns SQED with two electrons, depicted as a quiver in Fig. \ref{fig:SQED}. We will initially set both electrons' masses to 0. The Hilbert series of the theory is

\begin{multline}
\mathrm{HS}(t) = 1 + t( z +1 + \frac{1}{z} ) + t^2( z^2 + z + 1 + \frac{1}{z} + \frac{1}{z^2} ) + O(t^3)\\= 1 + t( w^2 +1 + w^{-2} ) + t^2( w^4 + w^2 + 1 + w^{-2} + w^{-4} ) + O(t^3) \\= 1 + t \lbrack 2 \rbrack + t^2 \lbrack 4 \rbrack + O(t^3)
\end{multline}
where $z\mapsto w^2$ cast it into a manifest sum of $\sl{2}$ characters $\lbrack n \rbrack = w^n + w^{n-2} + \dots + \frac{1}{w^n}$.

The series identifies a generator -- call it $N$ -- transforming in the (co)adjoint representation $\lbrack 2 \rbrack$. If the ring were freely generated then we would see a singlet $\lbrack 0 \rbrack$ and a tensor transforming in $\lbrack 4 \rbrack$ at quadratic order, but the singlet is absent. Hence there must be a quadratic singlet relation, which can only take the form $A \det N + B\ \mathrm{Tr} (N^2)=0$ for some $A,B$; a quick calculation shows that every generic choice of $A,B$ is equivalent\footnote{Exceptions such as $A=B=0$ would reduce the relation to $0=0$ and we can disregard them because the Hilbert series indicates there \emph{is} a non-trivial scalar relation.}. The relation can also be written as 

\be N^2=0,\label{eq:sl2adjrel}\ee
which identifies the space of $N$, ie. the Coulomb branch of this theory, as a nilpotent orbit of $\mathfrak{sl}(2)$.

This is a good result but some information is lost. There are three operators in $N$, but what \emph{are} they physically? How do they assemble into the matrix realisation of $N$? How should we physically interpret the relation $N^2=0$? If we set electrons' (complex) masses to $M$, would the relation change to $\mathrm{Tr} (N^2) = M^2$? Hilbert series can help with some of these questions but they are not the most suitable tools. 

Let's explore this problem using the algebraic construction of the chiral ring pioneered in \cite{Bullimore_The_2017}. This approach has several virtues: it is directly connected to physics and very cleanly handles complex mass deformations of the theory. However the Coulomb branch isometry remains hidden.

The ring is generated by two monopole operators $u^\pm$ and one scalar operator $\varphi$ subject to the relation

\be
u^+ u^- = -(\varphi - M_1)(\varphi - M_2) \label{eq:sl2rel}
\ee
where the $M_i$ are complex masses of electrons. It is important to note that this relation comes ``for free" from the definition of the chiral ring provided by \cite{Bullimore_The_2017}. This is a particularly simple example. There are no generators beyond $u^\pm$ and $\varphi$ and no relations beyond \ref{eq:sl2rel}. In other words, this is our chiral ring, but it is not immediately obvious that it describes (a deformation of) a nilpotent orbit of $\sl{2}$.

We want to develop a synthetic approach which adapts an important result of \cite{Bullimore_The_2017}: the Coulomb branch, being Hyperk\"{a}hler, has a moment map transforming in the coadjoint representation of $\sl{2}$ and specifically given by



\be
\mu = \left( \begin{matrix} \varphi - \frac{M_1}{2} -\frac{M_2}{2} & u^- \\ u^+ &-\varphi+ \frac{M_1}{2} + \frac{M_2}{2} \end{matrix} \right)
\ee

Recall that the adjoint and coadjoint representations of $\sl{2}$ are isomorphic and the Hilbert series has no way of distinguishing between them, so $N$ may in fact be a coadjoint generator. We will see that it is most naturally expanded in the coadjoint representation's basis as defined in this paper. 

$\mu$ also obeys the same relation as $N$ of \ref{eq:sl2adjrel}:

\be
\mu^2 = \left( \begin{matrix} \frac{1}{4}(M_1 + M_2 - 2 \varphi)^2 + u^+ u^- & 0 \\ 0 & \frac{1}{4}(M_1 + M_2 - 2 \varphi)^2 + u^+ u^- \end{matrix}\right) = \frac{1}{4} (M_1 - M_2)^2 \mathds{1}_{2\times 2}
\ee
where we used \ref{eq:sl2rel} to simplify some quadratic expressions. Note that when the masses are taken to 0 -- that is, precisely in the case considered using Hilbert series -- the equation reduces to $\mu^2=0$.

Several features of this result are noteworthy:

\begin{itemize}
	\item The matrix $\mu$ is traceless and hence belongs to  $\mathfrak{sl}(2)$ (or $\mathfrak{sl}(2)^*$) -- but is valued in the chiral ring $R$ rather than $\co$. The operator counting approach implied the existence of a coadjoint matrix $N$ whose complex coefficients are constrained by relations. The synthetic approach defines $\mu$ as a ring-valued matrix and matrix relations are reinterpreted as consequences of chiral ring relations which can be fully specified prior to embedding into a matrix.
	\item $\sl{2}$ has a natural (co)adjoint action on $\mu$ and components of $\mu$ generate the chiral ring -- so $\mu=N$.
	\item The fact that there are no independent higher-order relations is assured by Hilbert series. 
	\item However, the Coulomb branch Hilbert series provides no way of fixing the coefficient on the complex-mass-deformed relation. 
\end{itemize}

All of the above generalises to all examples considered in this paper and helps illustrate some of the utility of our synthetic method.

\subsection{$\sl{3}$: A slightly more complicated example}

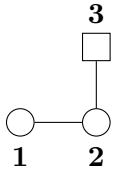
\begin{figure}[t]
	\centering
	\begin{tikzpicture}
		\node (g1) [gauge, label=below:{\textbf{1}}]{};
		\node (g2) [gauge, right of=g1, label=below:{\textbf{2}}]{};
		\node (f2) [flavor, above of=g2,label=above:{\textbf{3}}]{};
		\draw (g1)--(g2);
        \draw (g2)--(f2);
		\end{tikzpicture}
		\caption{Quiver with $\sl{3}$ isometry of the Coulomb branch}
		\label{fig:A2max}
\end{figure}

For the second example we pick the theory in Fig. \ref{fig:A2max}. Its gauge group is $U(1)\times U(2)$. Both gauge nodes are balanced so its Coulomb branch has an $A_2\simeq\sl{3}$ symmetry. We present its Hilbert series in terms of topological fugacities $z_1, z_2$ and $w_1, w_2$ related by

\be
w_i = \prod_j z_j^{\kappa^{-1}_{ij}}
\ee
where $\kappa_{ij}$ is the Cartan matrix
\be \bmat{cc} 2 & -1 \\ -1 & 2 \emat \ee
and we use the notation $\mathbf{\lbrack p_1, p_2 \rbrack}$ as shorthand for the $\sl{3}$ character with highest weight $\lbrack p_1, p_2 \rbrack$, eg.
\be
\mathbf{\lbrack 1,1 \rbrack} = w_1 w_2 + \frac{w_1^2}{w_2}+\frac{w_2^2}{w_1}+2+\frac{w_1}{w_2^2}+\frac{w_2}{w_1^2}+\frac{1}{w_1 w_2} \label{eq:sl3adj}
\ee

We call $w_i$, resp. $z_j$ \emph{fundamental weight}, resp. \emph{simple root fugacities} for reasons which will shortly become apparent.

This notation significantly simplifies the Hilbert series and manifests its nature as a class function:
\begin{multline}
\mathrm{HS}(t) = 1 + t (z_1 z_2 + z_1 + z_2 + 2 + \frac{1}{z_1} + \frac{1}{z_2} +\frac{1}{z_1 z_2}) + O(t^2) \\ = 1 + t \mathbf{\lbrack 1,1 \rbrack} + t^2 ( \mathbf{\lbrack 2,2 \rbrack} + \mathbf{\lbrack 1,1 \rbrack} ) + t^3 (\mathbf{\lbrack 3,3 \rbrack} + \mathbf{\lbrack 2,2 \rbrack} + \mathbf{\lbrack 3,0 \rbrack} + \mathbf{\lbrack 0,3 \rbrack}) + O(t^4)
\end{multline}

A closer look at the Hilbert series (to all orders) shows that the (massless) chiral ring is generated by a single $\sl{3}$ (co)adjoint tensor -- whose character appears in \ref{eq:sl3adj} -- subject to 

\be
\mathrm{Tr}\ N^2 = \mathrm{Tr}\ N^3 =0
\ee
which amounts to setting all eigenvalues to 0 and describes the maximal nilpotent orbit of $\sl{3}$.

The Hilbert series predicts 8 generators in total, two of which are linear Casimirs. Expressing $w_1^{p_1}w_2^{p_2}=\lbrack p_1, p_2 \rbrack$ and $z_1^{n_1} z_2^{n_2}=\rt{n_1,n_2}$, we observe the following correspondence to bare monopoles with magnetic charges $\vec{m}=(m_1;m_{2,1},m_{2,2})$:

\begin{align*}
	\lbrack 2,-1 \rbrack \leftrightarrow &\rt{1,0} \leftrightarrow \vec{m} = (1;0,0) \\
	\lbrack -1,2 \rbrack \leftrightarrow &\rt{0,1} \leftrightarrow \vec{m} = (0;1,0) \\
	\lbrack 1,1 \rbrack \leftrightarrow &\rt{1,1} \leftrightarrow \vec{m} = (1;1,0) \\
	\lbrack -2,1 \rbrack \leftrightarrow &\rt{-1,0} \leftrightarrow \vec{m} = (-1;0,0) \\
	\lbrack -1,2 \rbrack \leftrightarrow &\rt{0,-1} \leftrightarrow \vec{m} = (0;-1,0) \\
	\lbrack -1,-1 \rbrack \leftrightarrow &\rt{-1,-1} \leftrightarrow \vec{m} = (-1;-1,0) 
\end{align*}
It turns out that although the basis of fundamental weights is useful for pinning down the isometry and representation content, going back to $z_i$, or the basis of simple roots, is more physically transparent so we will keep working in that basis. 

We can now construct explicit generators and will label them as follows: generating monopole operators are indexed by corresponding roots, ie $\v{n_1,n_2}$, and linear Casimirs $\Phi$ carry the index of their gauge node, ie. $\Phi_i$. \cite{Bullimore_The_2017} provides a recipe to construct them in terms of auxiliary gauge-dependent \emph{abelianised} fields $\upm{1}, \p{1}, \upm{2,1}, \upm{2,2}, \p{2,1}$ and $\p{2,2}$:

\begin{align*}
\v{1,0} &= \up{1} \\
\v{0,1} &= \up{2,1}+\up{2,2}\\
\v{1,1} &= \frac{\up{1}\up{2,1}}{\p{1}-\p{2,1}}+\frac{\up{1}\up{2,2}}{\p{1}-\p{2,2}} \\
\v{-1,0} &= \um{1} \\
\v{0,-1} &= \um{2,1}+\um{2,2}\\
\v{-1,-1} &= \frac{\um{1}\um{2,1}}{\p{1}-\p{2,1}}+\frac{\um{1}\um{2,2}}{\p{1}-\p{2,2}}\\
\Phi_{1} &= \p{1}\\
\Phi_{2} &= \p{2,1}+\p{2,2}
\end{align*}

The algebraic construction also posits a set of relations:

\begin{align*}
\up{1}\um{1} &= - (\p{1}-\p{2,1})(\p{1}-\p{2,2}) \\
\up{2,1}\um{2,1} &= - \frac{(\p{2,1}-\p{1})(\p{2,1}-\m{2,1})(\p{2,1}-\m{2,2})(\p{2,1}-\m{2,3})}{(\p{2,1}-\p{2,2})^2} \\
\up{2,2}\um{2,2} &= - \frac{(\p{2,2}-\p{1})(\p{2,2}-\m{2,1})(\p{2,2}-\m{2,2})(\p{2,2}-\m{2,3})}{(\p{2,1}-\p{2,2})^2}	
\end{align*}

There are several structural features to point out. Firstly, operators such as $\p{2,1}$ and $\p{2,2}$ are gauge-dependent quantities; in fact, the Weyl group of $U(2)$ transforms one into the other. Their sum $\Phi_2 = \p{2,1}+\p{2,2}$, however, is gauge-invariant, as would be eg. $\p{2,1}\p{2,2}$. We will always reserve $\varphi$, resp. $\Phi$, for gauge-dependent, resp. gauge-independent manifestations of the scalar superpartners of gauge bosons and $\varphi_{i,a}$ will refer to the $a$-th gauge-dependent (\emph{abelianised}) scalar superpartner of the gauge bosons associated to the $i$-th node. 

Secondly, complex mass parameters $M_{i,p}$, again labelled as being the $p$-th mass on the $i$-th node, enter relations in a similar way to complex scalars $\varphi$. This is because complex masses can be interpreted as forming background supermultiplets with analogous coupling rules.

Thirdly, monopole operators $\v{\pm 1,\pm 1}$ have a curious structure of rational functions (and also the property of gauge-invariance-by-averaging which was just mentioned). The nature of such operators is, in our experience, a common source of confusion. One could think of e.g. $\up{1}\up{2,1}/(\p{1}-\p{2,1})$ as a new abstract ring element along with the relation  

\be
\frac{\up{1}\up{2,1}}{(\p{1}-\p{2,1})} (\p{1}-\p{2,1})=\up{1}\up{2,1} \label{eq:extmonorel}.
\ee
The chiral ring is still specifically a \emph{ring} and division is not in general defined as a valid operation.

Fourthly, the theory's chiral ring includes the quadratic Casimir operator $\p{2,1}\p{2,2}$ – in fact it's already present in the UV description. It is easy to check that

\be
\p{2,1}\p{2,2} = -\Phi_1(\Phi_1 - \Phi_2) - \v{1,0}\v{-1,0} \label{eq:quadrcasi}
\ee
Our method does not provide an algorithmic recipe for deriving this relation but its existence is ensured.

Finally, relations are given in terms of the abelianised and hence gauge-dependent fields. But the Coulomb branch only has directions corresponding to gauge-independent operators. So we would like to find gauge-independent relations to complement them; indeed, they should be exactly the relations predicted by Hilbert series. Our synthetic method can determine them.

The prescription for the coadjoint moment map (and the chiral ring generator) is

\be
N = \left( \begin{matrix} \Phi_1 - \frac{\m{2,1}+\m{2,2}+\m{2,3}}{3} & \v{-1,0} & -\v{-1,-1} \\ \v{1,0} &-\Phi_1 + \Phi_2- \frac{\m{2,1}+\m{2,2}+\m{2,3}}{3} & \v{0,-1} \\ -\v{1,1} & \v{0,1} & 2 \frac{\m{2,1}+\m{2,2}+\m{2,3}}{3} - \Phi_2 \end{matrix} \right)
\ee
and indeed, one easily finds that 

\begin{align}
	\mathrm{Tr}\ (N^2) &= \frac{2}{3} (\m{2,1}^2 + \m{2,2}^2 + \m{2,3}^2 - \m{2,1}\m{2,2} - \m{2,1}\m{2,3}-\m{2,2}\m{2,3}) \\
	\mathrm{Tr}\ (N^3) &= \frac{1}{9} (2\m{2,1}-\m{2,2}-\m{2,3}) (2\m{2,2}-\m{2,1}-\m{2,3}) (2\m{2,3}-\m{2,1}-\m{2,2})
\end{align}
both of which vanish in the massless limit. So we simultaneously derived gauge-invariant relations in the chiral ring and also generalised them for the case of massive quarks, demonstrating the advantages of the synthetic method over pure operator counting or algebraic construction.

\subsection{Construction of generators and gauge-dependent relations} \label{sec:monopoles}

All balanced quivers of type $A_n$ (of type $A$ with $n$ gauge nodes) and at least one gauge node of rank 1 share the same pattern of generators \cite{Hanany:2016gbz}. They always have $R$-symmetry spin 1 and include $n$ linear Casimirs originating from gauge scalars at the $n$ gauge nodes. The remaining generators are bare monopole operators labelled by their topological charges $\vec{q}=\rt{q_1,\dots,q_n}$\footnote{Topological charge vectors are written with angled brackets in anticipation of a thorough correspondence between their associated generating monopole operators and roots in the isometry algebra.} uniquely without any degeneracies. Every monopole generator exhibits the following pattern of charges: 

\be
\vec{q} = \rt{0,\dots,0,\pm 1,\dots,\pm 1,0,\dots,0},
\ee
or an uninterrupted string of $\pm 1$ padded by zeroes. The string of ones can stretch to each end so, for example, $\rt{1,1,1}$ is a valid charge vector of a monopole generator in an $A_3$ quiver. The choice of $+1$ or $-1$ must be made consistently in a given charge vector so no $A_3$ monopole generator carries the charge vector $\rt{1,-1,0}$ or other similarly ``mixed" charges. Such monopole operators still exist within the chiral ring but we do not count them among a canonical set of generators.

Overall we get $n^2 + n$ monopole operators and $n$ linear Casimirs which together generate the chiral ring. \cite{Bullimore_The_2017} provides a general prescription for these generators in terms of gauge-dependent quantities, or \emph{abelianised variables} as they are described in the original paper. The prescription was tested on several linear quivers in the original paper and succeeded when compared against known results. Principles behind the proposal have received further support in \cite{Bullimore_Vortices_2016,Assel_Ring_2017} which exploit quantum mechanics of vortices and string theory respectively. The chiral ring can be specified algorithmically:
\begin{itemize}
	\item Label each gauge node with an index $i \in \lbrace1,\dots,n \rbrace$ starting from the leftmost node. Let $r_i$ be the rank of the unitary group $U(r_i)$ at the gauge node $i$.
	\item Define the abelianised ring $R_\mathrm{abel}$.
		\begin{enumerate} 
			\item Any node with gauge group $U(r_i)$ and index $i$ gives rise to $3r_i$ abelianised variables: $u^+_{i,a}$, $u^-_{i,a}$ and $\varphi_{i,a}$, where $a$ runs from 1 to $r_i$. They physically correspond to directions in the moduli space of the fully broken gauge group $U(1)^{r_i}$. As an abelian theory it gives rise to $r_i$ different monopoles of charge $+1$ under the various $U(1)$ factors -- those would be the $u^+_{i,a}$ -- their counterparts with charges $-1$ -- the $u^-_{i,a}$ -- and complex scalars in the vector supermultiplet -- the $\varphi_{i,a}$. They are essentially eigenvalues of the adjoint-valued scalar superpartner of gauge bosons.
			\item We identify all topologically charged generators of the abelianised ring. Some of these operators carry no topological charge except $\pm 1$ at a single node $i$; we call such operators \emph{minimally charged} and they are already represented by $r_i$ operators $u^\pm_{i,a}$. The remaining monopole generators are topologically charged under several adjacent nodes and have to be constructed from the abelianised variables. They can be constructed in different (but equivalent) ways.
				\begin{itemize}
					\item \cite{Bullimore_The_2017} defines the Poisson bracket $\lbrace\cdot,\cdot\rbrace$ acting on the abelianised chiral ring; we reproduce it in \ref{eq:poiss}. An abelianised monopole charged under adjacent nodes $i$ and $i+1$ is given by \be \lbrace \upm{i,a},\upm{i+1,b}\rbrace \propto \frac{\upm{i,a}\upm{i+1,b}}{\p{i,a}-\p{i+1,b}} \ee
					with coefficient $\pm 1$. This can be extended by action of an adjacent node, eg. $\upm{i+2,c}$:\be \lbrace \frac{\upm{i,a}\upm{i+1,b}}{\p{i,a}-\p{i+1,b}},\upm{i+2,c}\rbrace \propto \frac{\upm{i,a}\upm{i+1,b}\upm{i+2,c}}{(\p{i,a}-\p{i+1,b})(\p{i+1,b}-\p{i+2,c})} \ee
					This operator can again be extended by the action of an adjacent node; the maximal operator ``stretches" between the leftmost and the rightmost nodes.
					\item Alternatively one can just give a general prescription for the non-minimally charged monopole generator. We will adopt this method and define a monopole charged $\pm 1$ under nodes $i, i+1, \dots, j-2, j-1$ as \be \upm{i:j,( a_i,\dots,a_{j-1})}=\frac{\upm{i,{a_i}}\cdots\upm{j-1,a_{j-1}}}{(\p{i,a_i}-\p{i+1,a_{i+1}})\cdots(\p{j-2,a_{j-2}}-\p{j-1,a_{j-1}})} \label{eq:abelmono}\ee
					In particular, $\upm{i,a}=\upm{i:i+1,(a)}$. Note that we selected the sign to be positive for all monopoles.
				\end{itemize}
			\item A flavor node of rank $s_i$ connected to the gauge node $i$ contributes complex mass parameters $M_{i,p}$, where $p$ runs from $1$ to $s_i$.
			\item Define $A(i)$ as the set of all nodes (resp. their indices) adjacent to node $i$; for most nodes $A(i)=\lbrace i-1,i+1\rbrace$.
			\item For each gauge node define two auxiliary polynomials: \be P_i(z) = \prod_{1\leq p \leq s_i} (z-M_{i,p}) \ee \be Q_i(z)= \prod_{1\leq a \leq r_i } (z-\p{i,a})\ee
			\item Abelianised variables are subject to relations\footnote{Note that these relations fix $R$-symmetry spin of bare abelianised monopoles $\upm{i,a}$ since topological charge conjugation should commute with $R$-symmetry and $\varphi$ have spin $1$.} 
				\be
				\up{i,a} \um{i,a} = - \frac{P_i(\p{i,a}) \prod_{j\in A(i)} Q_j(\p{i,a})}{\prod_{b\neq a} (\p{i,a}-\p{i,b})^2}
				\ee
				which can be repackaged as generators of the ideal
				\be
				I=\left< \up{i,a} \um{i,a} + \frac{P_i(\p{i,a}) \prod_{j\in A(i)} Q_j(\p{i,a})}{\prod_{b\neq a} (\p{i,a}-\p{i,b})^2} \right> \label{eq:relideal}
				\ee
			\item The abelianised ring $R_\mathrm{abel}$ is then a quotient of a polynomial ring freely generated by scalars and monopole generators:
				\be
				R_\mathrm{abel} = \co \lbrack \upm{i:j,( a_i,\dots,a_{j-1})},\p{i,a} \rbrack / I
				\ee 
				with $1\leq i < j \leq n+1$.
		\end{enumerate}
	\item  The overall gauge group of the quiver is $\mathcal{G}=\prod_i U(r_i)$. Its Weyl group is then $\mathcal{W}(\mathcal{G}) = \prod_{i} S_{r_i}$. $\mathcal{W}(\mathcal{G})$ has a natural action on the $u^\pm_{i,a}$ and $\varphi_{i,a}$: each $S_{r_i}$ permutes indices $a$ for a fixed $i$. The true, physical chiral ring $R$ can only include gauge-invariant operators; this is satisfied by restricting $R_\mathrm{abel}$ to $\mathcal{W}(\mathcal{G})$-invariant polynomials:
		\be
		R = R_\mathrm{abel}^{\mathcal{W}(\mathcal{G})} = \co \lbrack \upm{i:j,( a_i,\dots,a_{j-1})},\p{i,a} \rbrack^{\mathcal{W}(\mathcal{G})} / I
		\ee
		where $\upm{i:j,( a_i,\dots,a_{j-1})}$ are interpreted using \ref{eq:abelmono}\footnote{Some authors prefer treating $\upm{i:j,( a_i,\dots,a_{j-1})}$ as new abelian variables at the expense of loading the ideal $I$ with new relations analogous to \ref{eq:extmonorel}, which would effectively impose \ref{eq:abelmono}. This alternative viewpoint is arguably mathematically cleaner but we find ours more computationally convenient.} and indices are implicitly ranged over.
		
		Note that this construction manifestly includes Casimir invariants of scalar operators such as \ref{eq:quadrcasi} in the chiral ring and, if the Casimir invariant in question is not itself a generator, implies that it can be built up from other operators.
\end{itemize}

Several elements of $R$ are significant enough to deserve a name:

\begin{align}
	V^\pm_{i:j}&= \sum_{a,\dots,d} \upm{i:j,(a,\dots,d)} = \sum_{a,\dots,d} \frac{\upm{i,a}\cdots\upm{j-1,d}}{(\p{i,a}-\p{i+1,b})\cdots(\p{j-2,c}-\p{j-1,d})} \label{eq:monopolegens}\\ 
	\Phi_i &= \sum_{a} \p{i,a} \label{eq:scalargens}
\end{align}
Hilbert series computations for balanced type A quivers show that such operators form (at least some of) the generating set for $R$. It will also be helpful to repackage mass parameters into symmetric polynomials:

\begin{align}
M_i &= \sum_{p=1}^s M_{i,p}\\
\vec{M} &= (M_1,\dots,M_n) 	
\end{align}

\subsection{Chevalley-Serre basis} \label{sec:Abasis}

One of the goals of this paper is to assemble gauge-invariant generators of $R$ into an irreducible representation of the Coulomb branch symmetry. Provided the quiver is of the type described in Sec. \ref{sec:quiver}, all generators form a single \textit{coadjoint} representation of the symmetry. In the particular case of balanced linear quivers all $n^2 + 2n$ generators of the form \ref{eq:monopolegens} and \ref{eq:scalargens} assemble into a traceless $(n+1)\times(n+1)$ complex matrix and parametrise a subspace of all such matrices -- in particular a nilpotent orbit of $\sl{n+1}$. This section identifies an appropriate basis of the coadjoint representation $\sl{n+1}^*$ so that each basis vector corresponds to one chiral ring generator. We simply restate the choice of basis employed in \cite{Bullimore_The_2017} for type $A$ quivers but motivate the choice in a way that allows us to straightforwardly generalise to the novel case of type $D$\footnote{Analogous constructions also do the job for types $B$ and $C$.}.

In order to derive such a basis for the coadjoint representation we will first look for the basis of its dual, the corresponding \textit{adjoint} representation, which is equivalent to finding a particularly nice basis of the Lie algebra itself. 

We set off with a review of some basic facts about Lie algebras and declare our notation. A Lie algebra $\mathfrak{g}$ can be decomposed into two vector spaces

\be
\mathfrak{g}=\mathfrak{h}+\Phi,
\ee

where $\mathfrak{h}$, known as the \emph{Cartan subalgebra}, is the maximal commutative subalgebra generated by $n$ elements $h_i$, and $\Phi$ is its complement called the \emph{root space}. The root space can be partitioned into mutual eigenspaces of all $h_i$. Each subspace is one-dimensional and their generators are known as \emph{roots}. Crucially, the roots' eigenvalues under the action of $\lbrack h_i,\cdot\rbrack$, known as weights, are integer-valued. 

Complex simple Lie algebras are uniquely specified by their associated Dynkin diagrams up to isomorphisms. Conversely, given a Dynkin diagram one can reconstruct a Lie algebra isomorphism class. This is typically done using an appropriate \textit{Chevalley-Serre basis}. Given a Lie algebra $\mathfrak{g}$ described by a Dynkin diagram one can construct the corresponding $n\times n$ Cartan matrix $\kappa_{ij}$. The Chevalley-Serre basis is then generated (as a Lie algebra) by $n$ positive simple roots $\alpha^+_i$, $n$ negative simple roots $\alpha^-_i$ and $n$ generators $h_a$ of the commutative Cartan subalgebra $\mathfrak{h}$ together with a Lie bracket $\lbrack\cdot,\cdot\rbrack$ subject to relations

\begin{align}
	\label{eq:CW1}\lbrack h_a,h_b\rbrack &= 0\\
	\lbrack h_a,\alpha^\pm_j\rbrack &= \pm\kappa_{ja}\alpha^\pm_j \label{eq:cartansimple}\\
	\label{eq:CW3}\lbrack \alpha^+_i,\alpha^-_i\rbrack &= 2h_i\\
	\label{eq:CW4}\lbrack\alpha^\pm_i,\cdot\rbrack^{1-\kappa_{ij}}\alpha^\pm_j &= 0.
\end{align}
The final relation is called the \emph{Serre relation}.

The remaining elements of the Lie algebra $\mathfrak{g}$ are generated by repeated action of $\lbrack \alpha^\pm_i, \cdot \rbrack$. Note that this prescription only specifies a Lie algebra up to isomorphism.

Weight vectors $\vec{\lambda}^\alpha$ are defined by
\be
\lbrack h_a, \alpha \rbrack = \lambda^\alpha_a \alpha
\ee
and as mentioned above can be used to label generators of the root space $\Phi$.

Simple roots $\alpha^\pm_i$ are specifically represented by Cartan matrix row vectors $\lambda_a^{\alpha^\pm_i}=\pm(\vec{\kappa_i})_a=\pm\kappa_{ia}$. The basis in which integers $\lambda_a^{\alpha^\pm_i}$ are evaluated is called the \textit{basis of fundamental weights}. Although important in the theory of Lie algebras, it is less suitable for our purposes than the \textit{simple root basis}\footnote{There is a basis for the \emph{roots} -- in terms of fundamental weights or simple roots -- and a \emph{matrix basis} of $\sl{n+1}$ realising these same roots which we eventually use to construct a matrix realisation for the adjoint representation. The multiple uses of ``basis" should not be mutually confused.} which expands a root's eigenvalues in terms of eigenvalues of simple roots:

\be\label{eq:simplertbasis}
\vec{\lambda}^{\alpha^\pm} = \sum_{i=1}^n c_{i}^{\alpha^\pm} \vec{\lambda}^{\alpha^\pm_i} = \langle c^{\alpha^\pm}_1, \dots, c^{\alpha^\pm}_n \rangle
\ee

We use angled brackets to signify expansion in the simple root basis. The Jacobi identity implies that

 \be
 \lbrack h_i, \lbrack \alpha^\pm,\beta^\pm \rbrack \rbrack = (\lambda^{\alpha^\pm}+\lambda^{\beta^\pm})_a\lbrack \alpha^\pm,\beta^\pm \rbrack
 \ee
 
 This in particular implies that, since the Lie algebra is generated by brackets of simple roots, all $c_i^\pm$ are integers.
 
 Moreover, any positive (negative) root is constructed by finitely many bracket operations between positive (negative) roots, which implies we only need to expand $\vec{\lambda}$ in the eigenvalues of only positive (only negative) simple roots, as denoted by $\pm$ in superscripts of Eq. \ref{eq:simplertbasis}.
 
 One can easily convert vectors from the basis of fundamental weights to the simple root basis by multiplying with $\kappa^{-1}$ from the right:

\be
\lbrack \lambda^{\alpha^\pm}_1, \dots, \lambda^{\alpha^\pm}_n \rbrack (\kappa^{-1}) = \langle c^{\alpha^\pm}_1,\dots,c^{\alpha^\pm}_n\rangle
\ee

For a concrete example, consider the roots of $A_3$:

\be 
\begin{split}
\Phi_{A_3} = \{ &[1,0,1], [-1,1,1], [1,1,-1], [-1,2,-1], [2,-1,0],[0,-1,2],\\
&[0,1,-2],[-2,1,0],[1,-2,1],[-1,-1,1],[1,-1,-1],[-1,0,-1] \}
\end{split}
\ee

The numbers in square brackets state roots' fundamental weights. Multiplying on the right by the inverse of the Cartan matrix $\kappa^{-1}$ amounts to expressing a root in terms of the simple root basis (for which we use angled brackets). For example,

\begin{align*}
[1,0,1](\kappa^{-1} )&= \rt{1,1,1}\\
[1,1,-1](\kappa^{-1} )&= \rt{1,1,0}\\
[2,-1,0](\kappa^{-1} )&= \rt{1,0,0}\\
[0,-1,2](\kappa^{-1} )&= \rt{0,0,1}\\
[-1,-1,1](\kappa^{-1} )&= \rt{-1,-1,0}
\end{align*}

All roots of $A_n$ are given by unbroken strings of 1 or $-1$. Utility of the simple root basis lies partly in its exact correspondence with the set of topological charges exhibited by monopoles generators and partly in its description of the root's adjoint action. For example:

\begin{align}
\left[ \rt{1,0,0},\rt{0,1,0} \right]	 &\propto \rt{1,1,0}\\
\left[ \rt{1,1,0},\rt{0,0,1} \right]	 &\propto \rt{1,1,1}
\end{align}
Note that this mirrors the Poisson algebra defined on the chiral ring.

The precise coefficients, ie. \textit{structure constants}, are in this case $\pm 1$. While many relations between structure constants can be found, the constants are not uniquely fixed. Every choice produces a different (but isomorphic) algebra, so it makes more sense to speak of Chevalley-Serre \textit{bases}, each of which satisfies relations \ref{eq:CW1}–\ref{eq:CW4}. We will select the algebra which leaves monopole operators in their simplest form.

This section has so far treated elements of the Chevalley-Serre basis as abstract algebra elements (with a Lie bracket action) rather than concrete matrices (with the Lie bracket implemented through commutators). The remainder of this section is dedicated to construction of a concrete matrix realisation. In order to do this we introduce one final basis for roots: the \textit{orthonormal basis} given by $e_i - e_j$ where $e_i$ are the orthonormal basis vectors of $\C^{n+1}$. Simple roots are represented as

\be
\alpha^\pm_i \leftrightarrow \pm e_i \mp e_{i+1}
\ee

and brackets act by adding the orthonormal representatives, eg.

\be
\lbrack \alpha^+_1, \alpha^+_2 \rbrack \leftrightarrow e_1 - e_2 + e_2 - e_3 = e_1-e_3 \leftrightarrow \lbrack \alpha^+_2, \alpha^+_1 \rbrack.
\ee

This example demonstrates that the orthonormal representation loses some information -- namely the sign of the root's coefficient and hence the order in which two roots enter a Lie bracket -- but it still serves an important structural purpose. Since any root can be expressed in the simple root basis as an unbroken string of $\pm 1$, the $e_i - e_j$ cover and exhaust all roots. Each root is therefore labelled by two numbers, $i$ and $j$, with $i<j$ for positive roots and $j<i$ for negative roots. The orthonormal representation then provides a more compact labelling scheme for roots:
\begin{align}
\alpha^+_{i:j} \leftrightarrow  e_i-e_j\ (i<j)\\
\alpha^-_{i:j} \leftrightarrow  e_i-e_j\ (i>j) 
\end{align}
so in particular $\alpha^\pm_i=\alpha^\pm_{i:i+1}$. In words $\alpha^\pm_{i:j}$ is the root whose weight vector (in the simple root basis) consists of a string of $\pm 1$ starting at $i$ and terminating at $j-1$. 

It is now easy to guess that the matrix representatives of $\alpha^\pm_{i:j}$ is precisely the zero matrix with the $i,j$ or $j,i$ component changed to $\pm 1$ (according to the sign of the root and chosen convention for structure constants). Representatives of the Cartan subalgebra can be found by Eq. \ref{eq:CW3}. We pick the Chevalley-Serre basis given by 

\begin{align} 
(E_{i,j})_{ab} &= \delta_{ia} \delta_{jb} \\
\alpha_{i:j}^+ &= (-1)^{i-j+1} E_{i,j+1}\\
\alpha_{i:j}^- &= (-1)^{i-j+1} E_{j+1,i}\\
h_i &= E_{i,i}-E_{i+1,i+1}
\end{align}

The structure of alternating signs can already be seen in the following example of $\sl{4}$, where coefficients $c$ range over $\co$:

\be
\begin{split}
\mathrm{ad}(\sl{4})&=\left\{\sum_{\langle i,j,k \rangle \in \Phi} c_{\langle i,j,k \rangle} \langle i,j,k \rangle + \sum_{i=1}^3 c_{h_i} h_i\right\} \\
&= 
\left\{\left( \begin{matrix}
	c_{h_1} & c_{\langle 1,0,0 \rangle} & -c_{\langle 1,1,0 \rangle} & c_{\langle 1,1,1 \rangle}\\
	c_{\langle -1,0,0 \rangle} & -c_{h_1} + c_{h_2} & c_{\langle 0,1,0 \rangle} & -c_{\langle 0,1,1 \rangle}\\
	-c_{\langle -1,-1,0 \rangle} & c_{\langle 0,-1,0 \rangle} & -c_{h_2} + c_{h_3} & c_{\langle 0,0,1 \rangle}\\
	c_{\langle -1,-1,-1 \rangle} & -c_{\langle 0,-1,-1 \rangle} & c_{\langle 0,0,-1 \rangle} & -c_{h_3}
\end{matrix} \right)\right\} \\
&= \left\{\sum_{\substack{1\leq i<j\leq n\\s\in \lbrace +,-\rbrace}} c^s_{i:j} \alpha^s_{i:j} + \sum_{i=1}^3 c_{h_i} h_i\right\}\\
&= 
\left\{\left( \begin{matrix}
	c_{h_1} & c^+_{1:2} & -c^+_{1:3} & c^+_{1:4}\\
	c^-_{1:2} & -c_{h_1} + c_{h_2} & c^+_{2:3} & -c^+_{2:4}\\
	-c^-_{1:3} & c^-_{2:3} & -c_{h_2} + c_{h_3} & c^+_{3:4}\\
	c^-_{1:4} & -c^-_{2:4} & c^-_{3:4} & -c_{h_3}
\end{matrix} \right)\right\}
\end{split}
\ee


The final step is to identify the corresponding coadjoint basis which is dual to the adjoint basis with respect to the scalar product

\be
\langle X, Y \rangle = \tr{XY}.
\ee

Labelling elements of the Chevalley-Serre basis $X_m$ with the index $m$ ranging from $1$ to $\dim \mathfrak{g}$, we compute the matrix $C$

\be
C_{mn} = \tr{X_m X_n}.
\ee

Up to an overall multiplicative constant (the second order Dynkin index \cite{fuchs2003symmetries}), $C$ is precisely the Killing form. It is well known that the Killing form is non-degenerate and so $C$ can be inverted. We use it to define matrices

\be
X^*_m = \sum_p (C^{-1})_{mp} X_p \label{eq:dualroot}
\ee

satisfying the property

\be
\langle X^*_m, X_n \rangle = \tr{X^*_m X_n} = \sum_k (C^{-1})_{mp} \tr{X_p X_n} = (C^{-1} C)_{mn} = \delta_{mn} \label{eq:orthonormal}.
\ee

$X^*_m$ constitute the desired basis for the coadjoint representation of $\mathfrak{g}$ and dualisation $*:\mathfrak{g}\rightarrow\mathfrak{g}^*$ can be defined through linear extension of \ref{eq:dualroot}. 

For the Chevalley-Serre basis of type A one gets $\alpha^{\pm *}_{i:j}=\alpha^{\mp}_{i:j}$. On the other hand the Cartan subalgebra mixes in a non-trivial way, ie. elements of the Cartan subalgebra map to other elements in the subalgebra. $C|_\mathfrak{h} = \tr{H_i H_j}$, the restriction of the Killing form to $\mathfrak{h} \subset \mathfrak{g}$, is still non-degenerate, so we can define

\be
h^*_i = \sum_j (C|_\mathfrak{h})^{-1}_{ij} h_j
\ee


\subsection{Moment map}



The moment map of a symplectic space is a coadjoint-valued map, so we should be able to expand it in the basis \ref{eq:dualroot}. The coefficients will be precisely the vevs of the Coulomb branch operators of \ref{sec:monopoles}; in fact both the monopole generators and dual roots are labelled by unbroken strings of $\pm 1$ padded by zeroes and there are as many linear Casimirs as there are generators of the Cartan subalgebra, although here the correspondence is marginally more involved. 

The symplectic structure of the Coulomb branch gives rise to the Poisson bracket on operators \ref{eq:poiss}, which is closely related to the moment map and described by its action on the abelianised variables in \cite{Bullimore_The_2017}:

\begin{align}
\begin{split}
\pois{\p{i,a},\upm{i,a}} &= \pm\upm{i,a}\\
\pois{\up{i,a},\um{i,a}} &= \frac{\partial}{\partial \p{i,a}} \left[ \frac{P_i(\p{i,a}) \prod_{j\in A_i} Q_j(\p{i,a})}{\prod_{b\neq a}(\p{i,a}-\p{i,b})^2} \right]\\
\pois{\upm{i,a},\upm{j,b}} &= \pm \kappa_{ij}\frac{\upm{i,a}\upm{j,b}}{\p{i,a}-\p{j,b}}
\end{split}\label{eq:poiss}
\end{align}
The remaining undetermined brackets vanish.

 In fact, one can think of the moment map $N$ as a homomorphism from the Lie algebra of the Coulomb branch symmetry to the Poisson algebra of operators. More explicitly, for all $X_m, X_n \in \mathfrak{g}$

\be
\tr{N \lbrack X_m, X_n \rbrack} = \lbrace \tr{N X_m}, \tr{N X_n} \rbrace \label{eq:liepoiss}.
\ee

Before assembling $N$ we should identify the operator analogue of $h_i$ in \ref{eq:cartansimple}. Our simple roots are represented by operators $V^\pm_{i:i+1}$ and one can easily check that 
\be
\lbrace \sum_k \kappa_{ik} \Phi_k -M_i, V^\pm_{j:j+1} \rbrace = \pm \sum_k \kappa_{ik} \delta_{jk} V^\pm_{j:j+1} = \pm \kappa_{ij} V^\pm_{j:j+1}
\ee
and less easily, but straightforwardly on concrete cases, that
\be
\lbrace V^+_{i:i+1}, V^-_{i:i+1} \rbrace = \sum_k \kappa_{ik} \Phi_k -M_i.
\ee

We can then define $H_i \equiv \sum_k \kappa_{ik} \Phi_k - M_i$\footnote{Note that $M_i$ can be viewed as a scalar component of a background vector supermultiplet associated to the flavor node adjacent to $i$ and that the definition of $H_i$ treats it on the same footing as scalar components of vector supermultiplets of \textit{gauge} nodes $j$ adjacent to $i$, for which $\kappa_{ij}=-1$.} and construct the coadjoint-valued moment map:


\be
\begin{split}
N(\vec{M}) &= \sum_{\substack{1\leq i<j\leq n\\s\in \lbrace +,-\rbrace}} V^s_{i:j} \alpha^{s *}_{i:j} + \sum_{i=1}^n H_i h^*_i\\
&= \bmat{ccccc} 
\bar{\Phi}_1(\vec{M}) & V^-_{1:2} & -V^-_{1:3} & \cdots & (-1)^{n+1} V^-_{1:n+1} \\
V^+_{1:2} & -\bar{\Phi}_1(\vec{M})+\bar{\Phi}_2(\vec{M}) & V^-_{2:3} & \cdots & (-1)^{n}V^-_{2:n+1} \\
-V^+_{1:3} & V^+_{2:3} & -\bar{\Phi}_2(\vec{M})+\bar{\Phi}_3(\vec{M}) & \cdots & (-1)^{n-1}V^-_{3:n+1}\\
\hdots & \hdots & \hdots & \ddots & \vdots \\
(-1)^{n+1} V^+_{1:n+1} & (-1)^{n} V^+_{2:n+1} & (-1)^{n-1}V^+_{3:n+1} & \cdots & -\bar{\Phi}_n(\vec{M})
\emat
\end{split}
\label{eq:Amomentmap}
\ee
where $\bar{\Phi}_i(\vec{M})=(C^{-1}\kappa\Phi)_i - (C^{-1} \vec{M})_i$\footnote{$C^{-1}\kappa=\mathds{1}$ for type $A$ and $\frac{1}{2}\mathds{1}$ for type $D$, respectively, given our choices of bases.}. The homomorphism \ref{eq:liepoiss} follows from the definition of $N$ and \ref{eq:orthonormal}.

Hilbert series then predict that components of $N(\vec{0})$\footnote{We treat the complex masses $\vec{M}$ as \emph{parameters} of the theory rather than new moduli. Then $C^{-1} \vec{M}$ is just a vector of complex numbers and components of $N(\vec{0})$ are straightforwardly generated as shifts of components of $N(\vec{M})$ by constant numbers and vice versa, so the two generating sets are equivalent.} will generate the Coulomb branch chiral ring $R$:

\be
R = \co \lbrack N_{ij}(\vec{0}) \rbrack / I \label{eq:Agaugeinvchiralgens}
\ee  
where $I$ is the ideal of gauge-dependent relations as defined in \ref{eq:relideal}. 

This claim is already non-trivial (and was made in \cite{Bullimore_The_2017} for cases of type A). To see this note that as a gauge-invariant operator, the Casimir invariant $\sum_{1<a<b<r_i}\p{i,a}\p{i,b}$ can be found in the chiral ring. It should be possible to express it in terms of ring generators $N_{ij}(\vec{0})$ but that clearly cannot be done without invoking some relations in $I$ and we would like a guarantee that those relations are sufficient for this purpose. 

However, one should expect such a guarantee on theoretical grounds. On the one hand, the abelianisation approach manifestly includes all Casimir invariants of $\p{i,a}$. On the other hand, Casimir invariants of degree $d$ exhibit $R$-symmetry spin $d$ and all chiral rings considered in this paper are generated by operators of spin 1, as computed using Hilbert series methods. Therefore any Casimir invariants of degree greater than 1 must be equal to some combination of spin 1 operators. 

We are not aware of a generic formula for relations between Casimir invariants and moment map components but they can always be derived with a sensible ansatz: just try all linear combinations of generators with vanishing topological charges with the correct overall $R$-symmetry spin.

\subsection{Further examples}


Previous sections identify gauge-invariant generators of the chiral ring and lay the groundwork for generalisation to more general quivers. The current section concludes our investigation of quivers of type $A$ by expressing \ref{eq:Agaugeinvchiralgens} as a ring quotiented by an ideal of gauge-invariant relations.

\subsubsection{Minimal nilpotent orbit of $\sl{n+1}$} \label{sec:Amin}

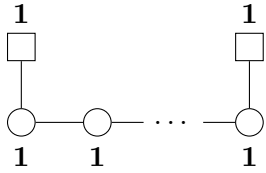
\begin{figure}[t]
	\centering
	\begin{tikzpicture}
		\node (g1) [gauge, label=below:{\textbf{1}}]{};
		\node (g2) [gauge, right of=g1, label=below:{\textbf{1}}]{};
        \node (dots) [right of=g2] {$\dots$};
		\node (gn) [gauge, right of=dots, label=below:{\textbf{1}}]{};
		\node (f1) [flavor, above of=g1,label=above:{\textbf{1}}]{};
		\node (fn) [flavor, above of=gn,label=above:{\textbf{1}}]{};
		\draw (g1)--(g2)--(dots)--(gn);
        \draw (gn)--(fn);
        \draw (g1)--(f1);
		\end{tikzpicture}
		\caption{A $3d \ \mathcal{N}=4$ quiver with $n$ gauge nodes. Its Coulomb branch is isomorphic to the minimal nilpotent orbit of $\sl{n+1}$ when the difference of complex mass parameters vanishes.}
		\label{fig:Amin}
\end{figure}

The Coulomb branch of the quiver in Fig. \ref{fig:Amin} is known from operator counting to be the minimal nilpotent orbit of $\sl{n+1}$ \cite{Hanany:2016gbz}, provided all mass parameters are set to 0. Then the Hilbert series identifies a single (co)adjoint generator $N$ subject to several relations transforming in particular representations. The only possible candidates are:

\ba
\mathrm{rank}\ N(\vec{0}) &< 2\\
\mathrm{Tr}N(\vec{0})^k &= 0 
\end{align}
where $k$ ranges from $1$ (trivially) to $n+1$; the second condition is equivalent to the vanishing of all eigenvalues of $N$.

One can now construct the chiral ring and the moment map \ref{eq:Amomentmap} to explicitly check that, in fact,

\be\label{eq:Aminrel1}
(N_i^a - \delta_i^a \frac{M_1-M_n}{n+1})(N_j^{b} - \delta_j^b \frac{M_1-M_n}{n+1}) - (a \leftrightarrow b) = 0
\ee

\be\label{eq:Aminrel2}
\mathrm{Tr}N^k - \frac{n(M_1-M_n)^k+(-n)^k(M_1-M_n)^k}{(n+1)^k}=0
\ee
where $N=N(\vec{M})$ and we redefined $M_i \eqdef M_{i,1}$ to reduce clutter.

This calculation is particularly tractable owing to the quiver's abelian gauge nodes and was partially done in \cite{Bullimore_The_2017}. Note that when complex mass parameters are set equal the equations reproduce predictions from Hilbert series. Moreover, the left hand sides of Eqs. \ref{eq:Aminrel1}-\ref{eq:Aminrel2} generate an ideal $J(\vec{M})$ of gauge-invariant operators. And in fact the Hilbert series implies that

\be
R = \co \lbrack N_{ij}(\vec{0}) \rbrack / I(\vec{M}) = \co \lbrack N_{ij}(\vec{0}) \rbrack / J(\vec{M})
\ee
$N_{ij}$ and $J(\vec{M})$ are both specified in terms of gauge-invariant operators, making good on our promise to define the chiral ring purely in terms of physically measurable moduli.

The space can be identified with $\mathrm{T}^* \mathbb{P}^{n}$ which is known to have a single deformation parameter, here the difference of masses.

\subsubsection{Maximal nilpotent orbit of $\sl{n+1}$} \label{sec:Amax}

\begin{figure}[t]
	\centering
	\begin{tikzpicture}
		\node (g1) [gauge, label=below:{\textbf{1}}]{};
		\node (g2) [gauge, right of=g1, label=below:{\textbf{2}}]{};
        \node (dots) [right of=g2] {$\dots$};
		\node (gn) [gauge, right of=dots, label=below:{\textbf{n}}]{};
		\node (fn) [flavor, above of=gn,label=above:{\textbf{n+1}}]{};
		\draw (g1)--(g2)--(dots)--(gn);
        \draw (gn)--(fn);
		\end{tikzpicture}
		\caption{The Coulomb branch of this quiver is isomorphic to the maximal nilpotent orbit of $\sl{n+1}$ when the difference of mass parameters vanishes.}
		\label{fig:Amax}
\end{figure}
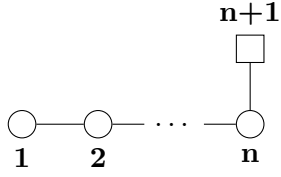

Coulomb branches of quivers depicted in Fig. \ref{fig:Amax} are isomorphic to maximal nilpotent orbits of $\sl{n+1}$ \cite{Hanany:2016gbz}. Hilbert series show that their chiral rings are again generated by the (co)adjoint generator $N$ defined by \ref{eq:Amomentmap}. The (massless) relations are known to be

\be
\mathrm{Tr}N(\vec{0})^k=0
\ee
for $1\leq k\leq n+1$.

Calculating complex-mass-deformed relations for general $n$ proves much more challenging than for minimal nilpotent orbits but numerical calculations at low enough $n$ are viable. It suffices to replace $N(\vec{0})\mapsto N(\vec{M})$ and straightforwardly evaluate\footnote{Complex masses were relabelled $M_{n,i}\rightarrow M_{i}$ for cleaner presentation.}:

\begin{itemize}
	\item 	$n=1$: \be \mathrm{Tr}N(\vec{M})^2 = \frac{1}{2} (M_{1} - M_{2})^2 \ee
	\item	$n=2$: \begin{align} \mathrm{Tr}N(\vec{M})^2 &= \frac{2}{3}(M_{1}^2+M_{2}^2+M_{3}^2 - M_{1}M_{2} - M_{1}M_{3} - M_{2}M_{3}) \\ \mathrm{Tr}N(\vec{M})^3 &= -\frac{1}{9}(-2M_{1}+M_{2}+M_{3})(M_{1}-2M_{2}+M_{3}) (M_{1} + M_{2}- 2 M_{3})\end{align}
	\item	$n=3$: \begin{align}\begin{split} \mathrm{Tr}N(\vec{M})^2 = \frac{1}{4}(&3M_{1}+3M_{2}+3M_{3}+3M_{4} - 2 M_{1}M_{2}-2M_{1}M_{3}-2M_{1}M_{4}\\&-2M_{2}M_{3}-2M_{2}M_{4}-2M_{3}M_{4}) \end{split} \\ \begin{split} \mathrm{Tr}N(\vec{M})^3 = \frac{3}{8} (&M_{1}^3+M_{2}^3+M_{3}^3+M_{4}^3-M_{1}^2M_{2}-M_{1}^2M_{3}-M_{1}^2M_{4}\\&-M_{2}^2M_{1}-M_{2}^2M_{3}-M_{2}^2M_{4}-M_{3}^2M_{1}-M_{3}^2M_{2}-M_{3}^2M_{4}\\&-M_{4}^2M_{1}-M_{4}^2M_{2}-M_{4}^2M_{3}+2M_{1}M_{2}M_{3}+2M_{1}M_{2}M_{4}\\&+2M_{1}M_{3}M_{4}+2M_{2}M_{3}M_{4})\end{split} \\ \begin{split}\mathrm{Tr}N(\vec{M})^4=\frac{1}{64}(&21M_{1}^4+21M_{2}^4+21M_{3}^4+21M_{4}^4-28M_{1}^3M_{2}-28M_{1}^3M_{3}\\&-28M_{1}^3M_{4}-28M_{2}^3M_{1}-28M_{2}^3M_{3}-28M_{2}^3M_{4}-28M_{3}^3M_{1}\\&-28M_{3}^3M_{2}-28M_{3}^3M_{4}-28M_{4}^3M_{1}-28M_{4}^3M_{2}-28M_{4}^3M_{3}\\&+30M_{1}^2M_{2}^2+30M_{1}^2M_{3}^2+30M_{1}^2M_{4}^2+30M_{2}^2M_{3}^2+30M_{2}^2M_{4}^2\\&+30M_{3}^2M_{4}^2+12M_{1}^2M_{2}M_{3}+12M_{1}M_{2}^2M_{3}+12M_{1}M_{2}M_{3}^2\\&+12M_{1}^2M_{2}M_{4}+12M_{1}M_{2}^2M_{4}+12M_{1}M_{2}M_{4}^2\\&+12M_{1}^2M_{3}M_{4}+12M_{1}M_{3}^2M_{4}+12M_{1}M_{3}M_{4}^2\\&+12M_{2}^2M_{3}M_{4}+12M_{2}M_{3}^2M_{4}+12M_{2}M_{3}M_{4}^2\\&+72M_{1}M_{2}M_{3}M_{4})\end{split}\end{align}
\end{itemize}


These relations are necessary and sufficient, as can be seen in their theories' Hilbert series.

\section{Type $D$} \label{sec:Dtype}
\subsection{$\mathfrak{so}(8)$: An example}

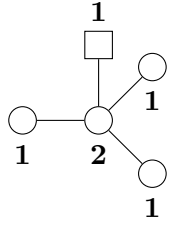
\begin{figure}[t]
	\centering
	\begin{tikzpicture}
		\node (g1) [gauge, label=below:{\textbf{1}}]{};
		\node (g2) [gauge, right of=g1, label=below:{\textbf{2}}]{};
		\node (g3) [gauge, above right of=g2, label=below:{\textbf{1}}]{};
		\node (g4) [gauge, below right of=g2, label=below:{\textbf{1}}]{};
		\node (f2) [flavor, above of=g2,label=above:{\textbf{1}}]{};
		\draw (g1)--(g2)--(g3);
		\draw (g2)--(g4);
        \draw (g2)--(f2);
		\end{tikzpicture}
		\caption{The Coulomb branch of this quiver is isomorphic to the $\so{8}$ minimal nilpotent orbit.}
		\label{fig:D4min}
\end{figure}

The synthetic method extends to balanced quivers of type $D$ and height 2 which we demonstrate on one of the simplest examples. The quiver in question, pictured in Fig. \ref{fig:D4min}, is shaped as the Dynkin diagram of $D_4$, suggesting $\so{8}$ isometry on the Coulomb branch. Its Hilbert series shows that the chiral ring is generated by 28 generators assembled into the (co)adjoint representation $N$ of $\so{8}$ \cite{Hanany:2016gbz}. The (massless) relations can also be identified through operator counting:

\begin{align}
N(\vec{0})^2	&=0 \label{eq:D4min1}\\
N(\vec{0})_{\lbrack ij}N(\vec{0})_{kl \rbrack} &= 0 \label{eq:D4min2}
\end{align}

The operators in $N$ correspond to 4 generators of the Cartan subalgebra, 12 positive roots and their 12 negative root counterparts. As expressed in the simple root basis, the positive roots are:

\be
\begin{split}
\Phi^+=\lbrace &\rt{1,0,0,0}, \rt{0,1,0,0}, \rt{0,0,1,0}, \rt{0,0,0,1}, \rt{1,1,0,0}, \rt{0,1,1,0}, \\ &\rt{0,1,0,1}, \rt{1,1,1,0}, \rt{1,1,0,1}, \rt{0,1,1,1}, \rt{1,1,1,1}, \rt{1,2,1,1} \rbrace
\end{split}
\ee

Roots label monopole operators by specifying charges at appropriate nodes: the first integer gives the topological charge under the leftmost node, followed by topological charges at the central, top right and finally bottom right node. Each node also contributes a topologically uncharged linear Casimir corresponding to the generator of the Cartan subalgebra $\mathfrak{h}\subset\so{8}$ carrying the same label. The fully assembled coadjoint generator – again playing the role of the moment map to the theory's Coulomb branch – is


\be
N(\vec{M})= \bmat{cccc} 
 \mathbf{J} \bar{\Phi}_1 & \mathbf{\bar{D}}^{\rt{1,0,0,0}}_{\rt{1,2,1,1}} &\mathbf{\bar{D}}^{\rt{1,1,0,0}}_{\rt{1,1,1,1}} &\mathbf{\bar{D}}^{\rt{1,1,1,0}}_{\rt{1,1,0,1}} \\
 -(\mathbf{\bar{D}}^{\rt{1,0,0,0}}_{\rt{1,2,1,1}})^T & \mathbf{J} (-\bar{\Phi}_1+\bar{\Phi}_2) &\mathbf{\bar{D}}^{\rt{0,1,0,0}}_{\rt{0,1,1,1}} &\mathbf{\bar{D}}^{\rt{0,1,1,0}}_{\rt{0,1,0,1}}\\
 -(\mathbf{\bar{D}}^{\rt{1,1,0,0}}_{\rt{1,1,1,1}})^T & -(\mathbf{\bar{D}}^{\rt{0,1,0,0}}_{\rt{0,1,1,1}})^T & \mathbf{J} (-\bar{\Phi}_2+\bar{\Phi}_3+\bar{\Phi}_4) & \mathbf{\bar{D}}^{\rt{0,0,1,0}}_{\rt{0,0,0,1}}\\
 -(\mathbf{\bar{D}}^{\rt{1,1,1,0}}_{\rt{1,1,0,1}})^T & -(\mathbf{\bar{D}}^{\rt{0,1,1,0}}_{\rt{0,1,0,1}})^T & -(\mathbf{\bar{D}}^{\rt{0,0,1,0}}_{\rt{0,0,0,1}})^T & \mathbf{J} (-\bar{\Phi}_3+\bar{\Phi}_4)
 \emat \label{eq:D4momentmap}
\ee
where
\begin{align}
\mathbf{D}^\alpha_\beta &= \frac{1}{4} \bmat{cc}\im (V_\alpha+V_{-\alpha}+V_\beta+V_{-\beta}) & V_\alpha - V_{-\alpha} - V_\beta +V_{-\beta} \\ -V_\alpha + V_{-\alpha} - V_{\beta} + V_{-\beta} & \im(V_\alpha+V_{-\alpha} -V_{\beta} -V_{-\beta})  \emat \\
\mathbf{J} &=\bmat{cc} 0 & \im \\ -\im & 0 \emat	\\
\bar{\Phi}_i &= \frac{1}{2}\Phi_i - (C^{-1}\vec{M})_i
\end{align}

The $V_\alpha$ and $\Phi_i$ are gauge-invariant objects which can be expressed in terms of gauge-dependent abelianised variables; those are in turn defined just as in Sec. \ref{sec:monopoles}. The explicit expressions are:

\begin{align}
	\Phi_1&=\p{1}\\
	\Phi_2&=\p{2,1}+\p{2,2}\\
	\Phi_3&=\p{3}\\
	\Phi_4&=\p{4}\\
	\v{\pm 1,0,0,0}&=\upm{1}\\
	\v{0,\pm 1,0,0}&=\upm{2,1}+\upm{2,2}\\
	\v{0,0,\pm 1,0}&=\upm{3}\\
	\v{0,0,0,\pm 1}&=\upm{4}\\
	\v{\pm 1, \pm 1, 0,0}&=\frac{\upm{1}\upm{2,1}}{\p{1}-\p{2,1}} + \frac{\upm{1}\upm{2,2}}{\p{1}-\p{2,2}}\\
	\v{0,\pm 1, \pm 1,0}&= \frac{\upm{2,1}\upm{3}}{\p{2,1}-\p{3}} + \frac{\upm{2,2}\upm{3}}{\p{2,2}-\p{3}}\\
	\v{0, \pm 1, 0, \pm 1}&= \frac{\upm{2,1}\upm{4}}{\p{2,1}-\p{4}} + \frac{\upm{2,2}\upm{4}}{\p{2,2}-\p{4}}\\
	\v{\pm 1, \pm 1, \pm 1, 0}&= \frac{\upm{1}\upm{2,1}\upm{3}}{(\p{1}-\p{2,1})(\p{2,1}-\p{3})}+\frac{\upm{1}\upm{2,2}\upm{3}}{(\p{1}-\p{2,2})(\p{2,2}-\p{3})} \\
	\v{\pm 1, \pm 1, 0, \pm 1}&= \frac{\upm{1}\upm{2,1}\upm{4}}{(\p{1}-\p{2,1})(\p{2,1}-\p{4})}+\frac{\upm{1}\upm{2,2}\upm{4}}{(\p{1}-\p{2,2})(\p{2,2}-\p{4})}\\
	\v{0,\pm 1, \pm 1, \pm 1}&= \frac{\upm{2,1}\upm{3}\upm{4}}{(\p{2,1}-\p{3})(\p{2,1}-\p{4})}+\frac{\upm{2,2}\upm{3}\upm{4}}{(\p{2,2}-\p{3})(\p{2,1}-\p{4})}\\
	\v{\pm 1,\pm 1, \pm 1, \pm 1} &= \frac{\upm{1}\upm{2,1}\upm{3}\upm{4}}{(\p{1}-\p{2,1})(\p{2,1}-\p{3})(\p{2,1}-\p{4})} + \frac{\upm{1}\upm{2,2}\upm{3}\upm{4}}{(\p{1}-\p{2,2})(\p{2,2}-\p{3})(\p{2,2}-\p{4})} \\
	\v{\pm 1, \pm 2, \pm 1, \pm 1} &= \frac{(\p{2,1}-\p{2,2})^2\upm{1}\upm{2,1}\upm{2,2}\upm{3}\upm{4}}{(\p{1}-\p{2,1})(\p{1}-\p{2,2})(\p{2,1}-\p{3})(\p{2,2}-\p{3})(\p{2,1}-\p{4})(\p{2,2}-\p{4})} \label{eq:D4highestroot}
\end{align}
with \ref{eq:relideal} acting on abelianised variables as the ideal of relations. A simple exercise in computer-assisted algebra is sufficient to check that \ref{eq:D4min1} and \ref{eq:D4min2} are satisfied by $N(\vec{0})$ and further that the gauge-invariant relations still hold without modification for $N(\vec{M})$:
\begin{align}
N(\vec{M})^2	&=0 \label{eq:D4min1}\\
N(\vec{M})_{\lbrack ij}N(\vec{M})_{kl \rbrack} &= 0 \label{eq:D4min2}	
\end{align}

This is not to say that complex mass parameters have no effect at all on the Coulomb branch: they modify the generator $N(\vec{M})$ itself by shifting scalar operators. However, this effect can be fully removed by redefining scalar fields with the opposite shift. The algebraic structure of relations \ref{eq:D4min1} and \ref{eq:D4min2} is also preserved in this particular case. Consequently, complex mass \emph{physically reparametrises} rather than \emph{deforms} this Coulomb branch.

Note that for \ref{eq:D4min2} this is the only result consistent with preservation of Coulomb branch isometry under mass deformation since there are no $\so{8}$-invariant tensors which could stand on the right hand side of that particular relation. \ref{eq:D4min1} could have been deformed by $(\vec{M}\cdot\vec{M}) \mathds{1}_{2n}$.


\subsection{Charges of chiral ring generators} \label{sec:Dcharges}

If the $D$-type quiver is of height 2 the chiral ring is generated by spin $1$ operators assembled into the adjoint representation of $\so{2n}$. The generators again split into linear Casimirs, of which there is one per node, and bare monopole operators labelled by topological charges. In this section we gather our knowledge about the latter.

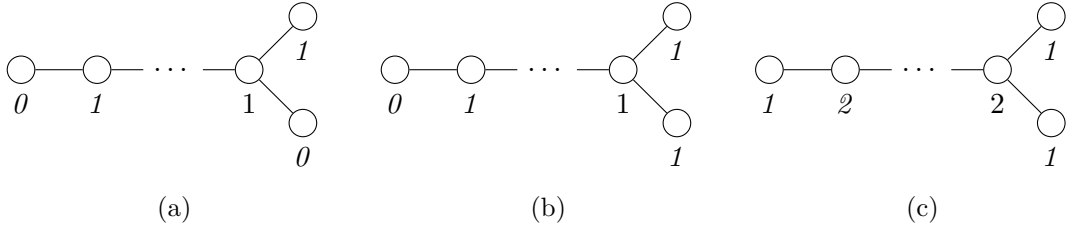
\begin{figure}[t]
	\centering
	\begin{subfigure}[t]{0.3\textwidth}
	\begin{tikzpicture}
		\node (g1) [gauge, label=below:{\textit{0}}]{};
		\node (g2) [gauge, right of=g1, label=below:{\textit{1}}]{};
		\node (gn-2) [gauge, right of=dots, label = below:{1}]{};
		\node (gn-1) [gauge, above right of=gn-2, label=below:{\textit{1}}]{};
		\node (gn) [gauge, below right of=gn-2, label=below:{\textit{0}}]{};
		\node (dots) [right of=g2]{$\dots$};
		\draw (g1)--(g2)--(dots)--(gn-2);
		\draw (gn-2)--(gn-1);
        \draw (gn-2)--(gn);
		\end{tikzpicture}
		\caption{}
		\label{fig:Dn1}
	\end{subfigure}	
~
	\begin{subfigure}[t]{0.3\textwidth}
	\begin{tikzpicture}
		\node (g1) [gauge, label=below:{\textit{0}}]{};
		\node (g2) [gauge, right of=g1, label=below:{\textit{1}}]{};
		\node (gn-2) [gauge, right of=dots, label = below:{1}]{};
		\node (gn-1) [gauge, above right of=gn-2, label=below:{\textit{1}}]{};
		\node (gn) [gauge, below right of=gn-2, label=below:{\textit{1}}]{};
		\node (dots) [right of=g2]{$\dots$};
		\draw (g1)--(g2)--(dots)--(gn-2);
		\draw (gn-2)--(gn-1);
        \draw (gn-2)--(gn);
		\end{tikzpicture}
		\caption{}
		\label{fig:Dn2}
	\end{subfigure}
~
	\begin{subfigure}[t]{0.3\textwidth}
	\begin{tikzpicture}
		\node (g1) [gauge, label=below:{\textit{1}}]{};
		\node (g2) [gauge, right of=g1, label=below:{\textit{2}}]{};
		\node (gn-2) [gauge, right of=dots, label = below:{2}]{};
		\node (gn-1) [gauge, above right of=gn-2, label=below:{\textit{1}}]{};
		\node (gn) [gauge, below right of=gn-2, label=below:{\textit{1}}]{};
		\node (dots) [right of=g2]{$\dots$};
		\draw (g1)--(g2)--(dots)--(gn-2);
		\draw (gn-2)--(gn-1);
        \draw (gn-2)--(gn);
		\end{tikzpicture}
		\caption{}
		\label{fig:Dn3}
	\end{subfigure}
		\caption{Numbers represent topological charges at each node.}
\end{figure}

Extensive sets of Hilbert series calculations \cite{Hanany:2016gbz} applied to these theories show that all monopole operators at $R$-symmetry spin 1 belong to one of two categories. The following classification identifies a monopole generator with a labelled quiver diagram whose flavour nodes and gauge rank information have been removed:

\begin{itemize}
	\item Unbroken (and linear) strings of either only $+1$ or only $-1$ stretching anywhere across the quiver -- see Fig \ref{fig:Dn1} for an example stretching all the way to the spinor node.
	\item Unbroken strings of $\pm 1$ (with uniform choice of sign) with charges $\pm 1$ on both rightmost (spinor) nodes -- see Fig \ref{fig:Dn2}. If both spinor nodes are turned on then a string of $\pm 2$ (with the same choice of sign as $\pm 1$) can be extended from the trivalent node arbitrarily far to the left, terminating with a string of $\pm 1$ which must have length at least 1 -- see Fig.\ref{fig:Dn3}.
\end{itemize}

It will prove convenient to arrange topological charges into linear vectors and we pick the usual convention, ie. the first $n-2$ entries describe charges on the linear segment from the first node to the trivalent node and the $n-1$-th, resp. $n$-th entries belong to the top right, resp. top bottom nodes.

\subsection{Construction of the chiral ring}

Construction of the chiral ring is closely analogous to that of Sec. \ref{sec:monopoles} with differences arising only with respect to monopoles whose topological charges stretch across multiple nodes.

The simplest and cleanest way to identify monopole operators is to utilise the symplectic structure defined in \cite{Bullimore_The_2017} and captured in the Poisson brackets of operators \ref{eq:poiss}.

Minimally charged (gauge-invariant) monopoles at node $i$ are defined as
\be
U^\pm_i=\sum_a \upm{i,a}
\ee
and we can use the action they induce along with the Poisson bracket, $\lbrace U^\pm_i,\cdot\rbrace$, to generate the entire set of bare monopole operators. The procedure is inductive on the sum of topological charges of a monopole, $q=\sum q_i$, where we treat positive and negative monopoles separately:

\begin{itemize}
	\item Restrict to positively charged monopole operators and take the first non-trivial case of $q=1$. These are the minimally charged monopoles and their description is given above.
	\item To get the expression for a positive monopole operator $V$ with topological charges $\vec{q}$ whose sum is $\sum_i q_i = q = r+1$ one can start by assuming the inductive hypothesis, that is, expressions are known for all bare monopole operators up to and including overall topological charge $r>1$. The classification of monopoles given in the previous section is enough to establish that there exists a monopole operator $V'$ with topological charges $\vec{r}$ such that $\sum_i r_i = r$ and $\vec{q}-\vec{r}$ is the usual unit vector $\vec{e}_i$. Then the monopole $V$ is obtained as follows:
	\be V=\pm\lbrace U^+_i,V'\rbrace \ee and the sign is chosen so that, when scalar fields in denominators are ordered ``lowest indices to the left, highest indices to the right" -- eg. in combinations $(\p{1}-\p{3})$ but not $(\p{4}-\p{2})$ -- the expressions are monic. This generates all positive monopoles.
	\item To generate negative monopoles merely replace positive abelianised monopole variables with their negative counterparts: $\up{i,a}\mapsto\um{i,a}$.
\end{itemize}

In the $\so{8}$ example the monopole operator with highest overall topological charge was obtained by
\be
\v{1,2,1,1} \propto \lbrace U^+_2,\v{1,1,1,1} \rbrace
\ee
and it is worth taking a look at the structure of $\ref{eq:D4highestroot}$ to see how this monopole operator arrives at overall $R$-symmetry spin 1.

\subsection{Chevalley-Serre basis}

The orthonormal basis for $D_n$ is exhausted by roots of the form $\pm e_i \mp e_j$ and $\pm e_i \pm e_j$ and the simple roots are in particular given by

\begin{equation}\label{eq:Dorthobasis}
\begin{split}
\alpha^\pm_i &\leftrightarrow \pm e_i \mp e_{i+1}	,\ 1\leq i \leq n-1 \\
\alpha^\pm_n &\leftrightarrow \pm e_{n-1} \pm e_{n}.
\end{split}
\end{equation}

The remaining roots are obtained through bracket products of simple roots. For example (using angled brackets to signify expansion in the simple root basis),

\begin{align}
	\lbrack \langle 1,1,0,0 \rangle,\langle 0,0,1,0 \rangle \rbrack &\propto \langle 1,1,1,0 \rangle\\
	\lbrack \langle 1,1,1,1 \rangle,\langle 0,1,0,0 \rangle \rbrack &\propto \langle 1,2,1,1 \rangle.
\end{align}

This corresponds to addition in the orthonormal basis:

\begin{align}
	\langle 1,1,1,0 \rangle &\leftrightarrow e_1 - e_4 = (e_1 - e_2) + (e_2 - e_3) + (e_3 - e_4)\\
	\langle 1,2,1,1 \rangle &\leftrightarrow e_1 + e_2 = (e_1 - e_2) + 2(e_2 - e_3) + (e_3 - e_4) + (e_3+e_4).
\end{align}

Whereas positive (negative) roots of $A_n$ corresponded to strings of 1 ($-1$) in the simple root basis, the corresponding structure is marginally more complicated for $D_n$ but it is exactly the same as that of monopole generators. We repeat (and very slightly fine-grain for the reader's convenience) the categorisation of roots from Sec \ref{sec:Dcharges}, augmenting it with information about the orthonormal basis:

\begin{enumerate}
	\item Unbroken strings of $\pm 1$ anywhere on the Dynkin diagram (see Fig. \ref{fig:Dn1}). They are the $\pm e_i \mp e_j$ and $\pm e_i \pm e_n$ in the orthogonal basis. \\
	\item $\pm 1$ on both spinor ($(n-1)$-th and $n$-th) nodes and an arbitrarily long string of $\pm 1$ towards the vector (first) node (see Fig. \ref{fig:Dn2}). They are the $\pm e_i \pm e_{n-1}$ in the orthogonal basis. \\
	\item $\pm 1$ on both spinor nodes, a string of $\pm 2$ starting at the $(n-2)$th node and terminating before the first node, continued by a string (of length at least 1) of $\pm 1$ toward the first node (see Fig. \ref{fig:Dn3}). They are the rest of the $\pm e_i \pm e_j$ in the orthogonal basis.
\end{enumerate}

We can therefore find two integers $i,j$ associated to each root, just as in the case of $A$ algebras. The complex Lie algebra of $D_n$, $\mathfrak{so}_\co (2n)$, acts linearly on the vector space $\co^{2n}$ and the adjoint representation therefore admits realisation as a $2n \times 2n$  antisymmetric matrix, which naturally breaks into $2\times 2$ blocks indexed precisely by $i,j=1,\dots,n$. Antisymmetry of matrices in $\mathfrak{so}_\co (2n)$ also relates the two off-diagonal $2\times 2$ blocks indexed by $i,j$ and $j,i$ (where $i\neq j$). This is schematically represented by the following matrix, which has zeroes everywhere apart from two $2\times 2$ blocks $\mathbf{D}$ sitting in the $(2i-1)$-th and $2i$-th row, $(2j-1)$-th and $2j$-th column and vice versa, modified by an overall constant dependent on the position of the $\mathbf{D}$ block within the larger matrix:

\be
\label{eq:Droots}
\mathscr{D}^{(ij)}=\begin{blockarray}{cccccc}
 \dots & 2i-1\ \&\ 2i & \dots & 2j-1\ \&\ 2j & \dots \\
   & \downarrow &  & \downarrow & \\
\begin{block}{(ccccc)c}
   &  &  &  &  &  \vdots \\
   &  &  & \im^{i-j+1}\mathbf{D}^{(ij)} &  & \leftarrow\ 2i-1\ \&\ 2i \\
   &  &  &  &  & \vdots \\
   & -(\im^{i-j+1})(\mathbf{D}^{(ij)})^T &  &  &  & \leftarrow\ 2j-1\ \&\ 2j \\
   &  &  &  &  &  \vdots \\
\end{block}
\end{blockarray}
\ee

Since the same indices $i$ and $j$ also label roots through the orthonormal basis, we should expect a correspondence between the two and indeed, each pair of off-diagonal blocks $\mathbf{D}$ contains precisely 4 complex degrees of freedom: just enough to represent all of $e_i-e_j$, $e_i + e_j$, $-e_i + e_j$ and $-e_i - e_j$ for $1\leq i<j\leq n$. Each root is represented by a slightly different $\mathbf{D}$ block, which we will denote $\mathbf{D}_{+-}$ for roots of the form $e_i-e_j$ ($i<j$), $\mathbf{D}_{++}$ for $e_i + e_j$ and $\mathbf{D}_{-+}$, $\mathbf{D}_{--}$ for their respective counterparts among negative roots. They are given by:

\begin{align*}
\mathbf{D}_{+-} &= \frac{\im}{2} \left(\begin{matrix}
	1 & \im\\
	-\im & 1
\end{matrix}\right)\\	
\mathbf{D}_{++} &= \frac{\im}{2} \left(\begin{matrix}
	1 & -\im\\
	-\im & -1
\end{matrix}\right)\\
\mathbf{D}_{-+} &= \frac{\im}{2} \left(\begin{matrix}
	1 & -\im\\
	\im & 1
\end{matrix}\right)\\	
\mathbf{D}_{--} &= \frac{\im}{2} \left(\begin{matrix}
	1 & \im\\
	\im & -1
\end{matrix}\right)
\end{align*}

The full block $\mathbf{D}$ is then a linear combination of the four matrices above,

\be
\label{eq:Dblock}\mathbf{D} = c_{+-}^{(ij)} \mathbf{D}_{+-} + c_{++}^{(ij)} \mathbf{D}_{++} + c_{-+}^{(ij)} \mathbf{D}_{-+} + c_{--}^{(ij)} \mathbf{D}_{--}.
\ee

Therefore the matrix realisation represents roots as

\begin{align}
\label{eq:Droot1}e_i - e_j &\leftrightarrow \mathscr{D}^{(ij)}|_{c^{(ij)}_{+-}=1} = \alpha^{(ij)}_{+-}\\
e_i + e_j &\leftrightarrow \mathscr{D}^{(ij)}|_{c^{(ij)}_{++}=1} = \alpha^{(ij)}_{++}\\
-e_i + e_j &\leftrightarrow \mathscr{D}^{(ij)}|_{c^{(ij)}_{-+}=1} = \alpha^{(ij)}_{-+}\\
\label{eq:Droot4}-e_i - e_j &\leftrightarrow \mathscr{D}^{(ij)}|_{c^{(ij)}_{--}=1} = \alpha^{(ij)}_{--}
\end{align}

where $1 \leq i < j \leq n$ and all other coefficients vanish.

All that remains is to define appropriate generators of the Cartan subalgebra, but that is easily achieved by invoking \ref{eq:CW3}. A Cartan subalgebra generator is given by

\be
\label{eq:Dcartan}h_i=\begin{blockarray}{cccccccc}
 & \dots  & $2i-1$\ \&\ $2i$ & $2i+1$\ \&\ $2i+2$ & \dots & & \\
   &  & \downarrow  & \downarrow & & & \\
\begin{block}{(cccccc)cc}
  \mathbf{0} &  &  &  &  &  & & \\
   & \ddots &  &  &  &  & & \vdots \\
   &  & \mathbf{H} & \mathbf{0} &  &  & \leftarrow\ & $2i-1$\ \&\ $2i$  \\
   &  & \mathbf{0} & -\mathbf{H} &  &  & \leftarrow\ & $2i+1$\ \&\ $2i+2$ \\
   &  &  &  & \ddots &  & & \vdots \\
   &  &  &  &  & \mathbf{0} & & \\
\end{block}
\end{blockarray}
\ee

for $i=1,\dots,n-1$, where

\be
\label{eq:Hblock}\mathbf{H}=\left(\begin{matrix}
	 0 & \im\\
	-\im & 0
\end{matrix}\right)
\ee

and the remaining entries of $h_i$ are zero. The final Cartan generator differs only very slightly from $h_{n-1}$, as one might expect:

\be
h_n=\begin{blockarray}{cccccc}
 & \dots  & $2n-3$\ \&\ $2n-2$ & $2n-1$\ \&\ $2n$ &  \\
   &  & \downarrow  & \downarrow &  \\
\begin{block}{(cccc)cc}
  \mathbf{0} &  &  &  & &  \\
   & \ddots &  &  &  & \vdots \\
   &  & \mathbf{H} & \mathbf{0} & \leftarrow\ & $2n-1$\ \&\ $2n$  \\
   &  & \mathbf{0} & \mathbf{H} & \leftarrow\ & $2n+1$\ \&\ $2n+2$ \\
\end{block}
\end{blockarray}.
\ee

The full adjoint representation is then realised as

\be
\mathrm{adj}(\so{2n})=\left\{\sum_{\substack{1\leq i < j \leq n \\ a,b\ \in \{ +, - \}}} c^{(ij)}_{ab} \alpha^{(ij)}_{ab}  + \sum_{1\leq i \leq n} c_{h_i} h_i\right\} \label{eq:Dadj}
\ee

where coefficients $c$ range over $\co$.

We were unable to find an earlier matrix realisation of the $\so{2n}$ Chevalley-Serre basis and had to construct it ourselves. Therefore as far as we can tell its form is an original result. We will gladly accept corrections and references to prior work.

As was the case with type $A$ Chevalley-Serre bases, we finish this section by identifying the basis of the coadjoint representation. The generalisation is completely straightforward. We define the dual of a root $X^*_m \equiv \sum_n (C^-1)_{mn}X_n$ through the inverse of the matrix

\be
C_{mn} = \tr{X_m X_n},
\ee
which is again proportional to the non-degenerate Killing form. As was the case with type $A$, positive roots are swapped with their negative counterparts, although now an overall rescaling factor is involved:

\be
\alpha^{(ij)*}_{ab}=\frac{1}{2}\alpha^{(ij)}_{(-a)(-b)}
\ee

There is no additional subtlety in the dualisation of the Cartan subalgebra, which again mixes non-trivially through the the restriction of the Killing form to $\mathfrak{h}$:

\be
h^*_i = \sum_j (C|_\mathfrak{h})^{-1}_{ij} h_j.
\ee


\subsection{Moment map}


All that remains to define the Coulomb branch moment map is to associate generators of the coadjoint basis with monopole and linear Casimir operators. 

\begin{itemize}
	\item For monopole operators use \ref{eq:Dorthobasis} to translate labels in the simple root basis into the orthonormal basis:
	\be V_\alpha \leftrightarrow V^{(ij)}_{ab} \ee where $a,b\in \lbrace +,-\rbrace$ and $1\leq i < j \leq n$ and pair them with the corresponding dual roots: \be \alpha^{(ij)*}_{ab} \leftrightarrow V^{(ij)}_{ab} \ee
	\item Linear Casimirs need to be suitably combined to reproduce Poisson brackets analogously to the case of type $A$; a mass shift is also allowed by the abelianised Poisson brackets: \be h^*_i \leftrightarrow H_i = \sum_j \kappa_{ij} \Phi_j - M_i  \ee
\end{itemize}

Putting everything together the moment map comes out as

\be
N = \sum_{\substack{1\leq i < j \leq n \\ a,b\ \in \{ +, - \}}} V^{(ij)}_{ab} \alpha^{(ij)*}_{ab}  + \sum_{1\leq i \leq n} H_i h^*_i
\ee

This prescription tends to lead to matrices which struggle to fit on a page so we refer to the case of $\so{8}$ in \ref{eq:D4momentmap} as an exemplar.

The moment map still generates the Lie algebra homomorphism \ref{eq:liepoiss}, albeit for a $D_n$ algebra.

\subsection{Further examples}
\subsubsection{Minimal nilpotent orbit of $\so{2n}$}

\begin{figure}[t]
	\centering
	\begin{tikzpicture}
		\node (g1) [gauge, label=below:{\textbf{1}}]{};
		\node (g2) [gauge, right of=g1, label=below:{\textbf{2}}]{};
		\node (dots) [right of=g2]{$\dots$};
		\node (gn-2) [gauge, right of=dots, label=below:{\textbf{2}}]{};
		\node (gn-1) [gauge, above right of=gn-2, label=below:{\textbf{1}}]{};
		\node (gn) [gauge, below right of=gn-2, label=below:{\textbf{1}}]{};
		\node (f2) [flavor, above of=g2,label=above:{\textbf{1}}]{};
		\draw (g1)--(g2)--(dots)--(gn-2);
		\draw (gn-2)--(gn-1);
		\draw (gn-2)--(gn);
        \draw (g2)--(f2);
		\end{tikzpicture}
		\caption{The Coulomb branch of this quiver is the minimal nilpotent orbit of $D_n$ where $n$ is the number of gauge nodes.}
		\label{fig:Dmin}
\end{figure}
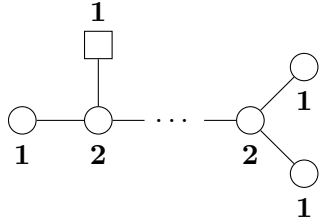

The $D_n$ analogue of quivers investigated in \ref{sec:Amin} is depicted in Fig. \ref{fig:Dmin}. Their Coulomb branches are the closures of minimal nilpotent orbits of $D_n$. The conditions on such an orbit are

\begin{align}
N(\vec{0})^2&=0\\
\mathrm{rank}\ N(\vec{0}) &<2	
\end{align}

and have been numerically verified for low values of $n$. The lack of a complex mass deformation in the minimal nilpotent orbit of $\so{8}$ generalises to minimal nilpotent orbits of $\so{2n}$ with $n>4$.

\subsubsection{Next-to-minimal nilpotent orbit of $\so{8}$}

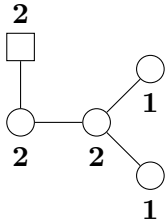
\begin{figure}[t]
	\centering
	\begin{tikzpicture}
		\node (g1) [gauge, label=below:{\textbf{2}}]{};
		\node (g2) [gauge, right of=g1, label=below:{\textbf{2}}]{};
		\node (g3) [gauge, above right of=g2, label=below:{\textbf{1}}]{};
		\node (g4) [gauge, below right of=g2, label=below:{\textbf{1}}]{};
		\node (f1) [flavor, above of=g1,label=above:{\textbf{2}}]{};
		\draw (g1)--(g2)--(g3);
		\draw (g2)--(g4);
        \draw (g1)--(f1);
		\end{tikzpicture}
		\caption{The Coulomb branch of this quiver is isomorphic to the $\so{8}$ next-to-minimal nilpotent orbit.}
		\label{fig:D4nexttomin}
\end{figure}

We provide one final example of $D_n$ nilpotent orbits, the next-to-minimal nilpotent orbit in Fig. \ref{fig:D4nexttomin}. The relations are known to be
\begin{align}
\mathrm{Tr} N(\vec{0})^2 &= 0 \\
N(\vec{0})_{\lbrack ij} N(\vec{0})_{kl \rbrack} &= 0	
\end{align}
and have been verified by our methods. Turning on masses leads to the related set of equations
\begin{align}
\mathrm{Tr} N(\vec{M})^2 &= \frac{1}{2} (M_{1,1}-M_{1,2})^2 \\
N(\vec{M})_{\lbrack ij} N(\vec{M})_{kl \rbrack} &= 0	
\end{align} 
The trace equation shows that this Coulomb branch has a complex mass deformation.

\section{Summary}

We aimed to demonstrate a certain kind of workflow for investigations of $3d$ $\mathcal{N}=4$ Coulomb branches:

\begin{enumerate}
	\item Calculate the Hilbert series and identify representations of generators and relations under the Coulomb branch isometry.
	\item Explicitly construct gauge-invariant monopole operators and scalar operators out of abelianised variables and attempt to assemble them into the aforementioned generator representations.
	\item Test gauge-invariant relations at the SCFT point and, if successful, turn on complex mass parameters to identify SUSY-preserving deformations of the Coulomb branch.
\end{enumerate}

While our examples only cover a narrow slice of available quiver theories we believe the general workflow fully generalises to many (all?) $3d$ $\mathcal{N}=4$ theories.

\section{Future developments} \label{sec:future}

This work develops several results of \cite{Bullimore_The_2017}, particularly its explicit and physically interpretable construction of the Coulomb branch moment map for many balanced unitary quivers of type $A$. We were able to extend our understanding to a subclass of type $D$ quivers.

Such results naturally call for further extension to quivers of types $B$, $C$, $E$, $F$ and $G$ and indeed we intend to carry out these investigations in the near future. $B$ and $C$ cases require the development of \emph{quiver folding}, a non-trivial procedure along the lines of \cite{Hanany:2018dvd}, which we hope to address in upcoming work. Since our method embeds gauge-operators into a matrix realisation of the isometry's coadjoint representation, it is most readily suited for cases in which the isometry is described by a classical algebra. Quivers of types $E$, $F$ and $G$ would require a different approach.

Increase in quiver height adds several new generators to the chiral ring of type $D$ quivers. It would be interesting to express them in terms of abelianised variables and construct their gauge-invariant relations. A similar phenomenon appears upon generalisation to quivers without a $U(1)$ node and our methods could provide a novel window into quiver subtractions of \cite{Cabrera:2018ann}.

We may also sacrifice balance. Quivers with one overbalanced node (excess greater than 0) were recently identified as relevant to the vacuum structure of five-dimensional supersymmetric theories. Such quivers' chiral rings are generated by a tensor in the coadjoint representation and additional tensors in another representation of the overall symmetry. We have also studied classes of ugly quivers in a so far unpublished research note written jointly with S. Cabrera.

Finally, it should be possible to extend our methods to orthosymplectic quivers but such a move would require a generalisation of the analysis in \cite{Bullimore_The_2017} along the lines of \cite{Assel:2018exy}.

\section{Acknowledgments}
The authors are grateful for fruitful discussions with Santiago Cabrera, Rudolph Kalveks, Benjamin Assel, Tudor Dimofte and Davide Gaiotto. D.M. is particularly indebted to Matthew Bullimore for his patient explanation of \cite{Bullimore_The_2017} which sparked this investigation and would like to thank Julius Grimminger for spotting several typos in an earlier version of this paper. A.H. is supported by STFC Consolidated Grant ST/J0003533/1, and EPSRC Programme Grant EP/K034456/1. D.M. is supported by STFC DTP research studentship grant ST/N504336/1.


\bibliographystyle{hieeetr}
\bibliography{ad_nilpotent_orbits}

\end{document}